\documentclass[twocolumn]{article}
\usepackage{mathtools}
\usepackage{graphicx}
\usepackage{color}
\usepackage{upgreek}
\usepackage{float}
\usepackage{txfonts}
\usepackage{longtable}
\usepackage[comma,super,sort&compress]{natbib}
\usepackage[colorlinks = true,
            linkcolor = blue,
            urlcolor  = blue,
            citecolor = blue,
            anchorcolor = blue]{hyperref}

\makeatother

\begin{document}

\title{Detection of two bright radio bursts from magnetar SGR~1935$+$2154}

\author{F.~Kirsten$\mathrm{^{*}}$\\
        Department of Space, Earth and Environment, Chalmers University of Technology \\ Onsala Space Observatory, 439 92, Onsala, Sweden \\
         \texttt{franz.kirsten@chalmers.se} 
         \and
        M.\,P.~Snelders \\
        Anton Pannekoek Institute for Astronomy, University of Amsterdam \\ Science Park 904, 1098 XH, Amsterdam, The Netherlands
        \and
        M.~Jenkins \\
        Anton Pannekoek Institute for Astronomy, University of Amsterdam \\ Science Park 904, 1098 XH, Amsterdam, The Netherlands
        \and
        K.~Nimmo \\
        Anton Pannekoek Institute for Astronomy, University of Amsterdam \\ Science Park 904, 1098 XH, Amsterdam, The Netherlands \\
        ASTRON, Netherlands Institute for Radio Astronomy \\ Oude Hoogeveensedijk 4, 7991 PD Dwingeloo, The Netherlands
        \and
        J.~van den Eijnden \\
        Anton Pannekoek Institute for Astronomy, University of Amsterdam \\ Science Park 904, 1098 XH, Amsterdam, The Netherlands
        \and
        J.\,W.\,T.~Hessels \\
        Anton Pannekoek Institute for Astronomy, University of Amsterdam \\ Science Park 904, 1098 XH, Amsterdam, The Netherlands \\
        ASTRON, Netherlands Institute for Radio Astronomy \\ Oude Hoogeveensedijk 4, 7991 PD Dwingeloo, The Netherlands
        \and
        M.\,P.~Gawro\'nski \\
        Institute of Astronomy, Faculty of Physics, Astronomy and Informatics, Nicolaus Copernicus University \\ Grudziadzka 5, 87-100 Toru\'n , Poland
        \and
        J.~Yang \\
        Department of Space, Earth and Environment, Chalmers University of Technology \\ Onsala Space Observatory, 439 92, Onsala, Sweden 
      }
          
\maketitle

\section{Abstract} 
   Fast radio bursts (FRBs) are millisecond-duration, bright radio signals (fluence $\mathrm{0.1 - 100\,Jy\,ms}$) emitted from extragalactic sources of unknown physical origin. The recent CHIME/FRB and STARE2 detection of an extremely bright (fluence $\sim$MJy$\,$ms) radio burst from the Galactic magnetar SGR~1935$+$2154 supports the hypothesis that (at least some) FRBs are emitted by magnetars at cosmological distances.
   In follow-up observations totalling 522.7$\,$hrs on source, we detect two bright radio bursts with fluences of $112\pm22\mathrm{\,Jy\,ms}$ and $24\pm5\mathrm{\,Jy\,ms}$, respectively. Both bursts appear affected by interstellar scattering and we measure significant linear and circular polarisation for the fainter burst.
   The bursts are separated in time by $\sim$1.4$\,$s, suggesting a non-Poissonian, clustered emission process -- similar to what has been seen in some repeating FRBs. Together with the burst reported by  CHIME/FRB and STARE2, as well as a much fainter burst seen by FAST (fluence 60$\mathrm{\,mJy\,ms}$), our observations demonstrate that SGR~1935$+$2154 can produce bursts with apparent energies spanning roughly seven orders of magnitude, and that the burst rate is comparable across this range. This raises the question of whether these four bursts arise from similar physical processes, and whether the FRB population distribution extends to very low energies ($\sim10^{30}\,$erg, isotropic equivalent).

\section{Introduction} 
Many different progenitor and emission models have been proposed to explain the FRB phenomenon\cite{platts_2019}, with one popular class of theories invoking neutron stars with exceptionally strong ($10^{14} - 10^{16}\,\mathrm{G}$) magnetic fields, commonly known as magnetars. Until now, the absence of multi-wavelength detections of prompt emission\cite{chen_2020ApJ...897..146C, 2017ApJ...846...80S} as well as the large distances to FRBs (\citep[FRB~180916.J0158+65  is the closest known, at $\sim$150$\,$Mpc][]{Marcote_2020}) have made it hard to study their broadband emission mechanism and local environments.  This limits the avenues to differentiate between competing models.  The localisation of very nearby (tens of Mpc) FRBs could help, as would the discovery of an FRB source, at kpc distances, in the Milky Way.

On 2020 April 28 a breakthrough was made when \citet{chime_magnetar_2020_arxiv} and \citet{bochenek_2020_arxiv} independently detected an extremely bright radio burst from the Galactic magnetar SGR~1935$+$2154, using the Canadian Hydrogen Intensity Mapping Experiment Fast Radio Burst Project (\citep[CHIME/FRB][]{chimefrb_2018ApJ...863...48C}) and the Survey for Transient Astronomical Radio Emission 2 (\citep[STARE2][]{stare2_bochenek_2020PASP..132c4202B}), respectively. The reported burst fluence was $\mathrm{1.5\,MJy\,ms}$ at 1.4$\,$GHz \citep{bochenek_2020_arxiv}, and the equivalent isotropic energy of the burst was approximately three orders of magnitude greater than any previously observed magnetar radio burst. The specific energy of the burst is similar to, although approximately 30 times less than, the specific energy of the faintest known FRB \citep{bochenek_2020_arxiv, Marcote_2020}.  These detections strongly suggest that at least some FRBs are produced by magnetars.  For this reason, this burst has been referred to as FRB~200428 in the literature.  While it is not conclusively established that this burst comes from the same physical process(es) as extragalactic FRBs, we will nonetheless use this nomenclature for the rest of this paper.

 Temporally coincident with the radio pulse, a bright, hard X-ray burst was detected independently by the Konus-Wind\cite{ridnaia_2020_arxiv}, INTEGRAL\cite{Mereghetti_integral}, AGILE\cite{AGILE_ATel_13686}, and Insight-HXMT\cite{li_insight_hxmt_simultaneous_CHIME_STARE2_Xray_burst_2020_arxiv} satellites. SGR~1935$+$2154 has been known to undergo periods of X-ray outbursts in 2014, 2015, and 2016, but simultaneous radio observations at these times did not produce any significant detections\cite{sgr_14_15_16_outbursts}.  The radio bursts from this most recent outburst are the first to be detected from this source, and the simultaneous radio/X-ray detection is a first for any Galactic magnetar (or FRB source) in general.

A few days after the announcement of FRB~200428,  \citet{fast_polarised_burst_zhang_2020ATel13699....1Z} used the Five-hundred-meter Aperture Spherical radio Telescope (\citep[FAST][]{Nan2011}) to detect a much fainter (fluence 60$\mathrm{\,mJy\,ms}$), highly linearly polarised burst from SGR~1935$+$2154.  Its polarisation properties are very similar to FRB~121102 \citep{michilli_2018_magnetoionic} and FRB~180916.J0158+65 \citep{chime_2019}.

The detection of more radio bursts from SGR~1935$+$2154, and a more detailed characterisation of its activity levels, can help understand whether it is genuinely an FRB source, with similar physical nature to the sources of (repeating) extragalactic FRBs.  Given the great brightness of FRB~200428, a coordinated campaign of small radio telescopes (25-m diameter)  with large on-sky time (hundreds of hours) can complement deeper, but shorter campaigns using larger radio telescopes.  Furthermore, the relatively narrow-band emission seen from some FRBs \citep{2019ApJ...876L..23H,2019ApJ...877L..19G,majid_2020} motivates a coordinated, multi-telescope campaign that spans a wide range of radio frequencies simultaneously.

\section{The data} 
Between 2020 April 29 and 2020 July 27 we observed SGR~1935$+$2154 for a total of 763.3$\,$hrs, which corresponds to 522.7$\,$hrs of on-source time, taking overlap between the participating stations into account. The stations involved were the 25-m single dish RT1
at Westerbork in the Netherlands, the 25-m and 20-m telescopes at Onsala Space Observatory (OSO) in Sweden, and the 32-m dish in Toru\'n, Poland (see Table \ref{tab:coverage} and Methods for details). All stations operated independently as single dishes,
recording 2-bit baseband data (circular polarisations) in VLBI Data Interchange Format
(\cite[VDIF][]{2010ivs..conf..192W}). The data from all four stations were processed and searched for bursts at OSO using a pipeline that was developed to search for FRBs in baseband data. In essence, the pipeline uses standard pulsar software (\cite[DSPSR][]{vanStraten_digifil_2011}) in combination with \href{https://sourceforge.net/projects/heimdall-astro/}{{\tt Heimdall}} and FETCH\cite{agarwal19_fetch_arxiv} to create channelised total intensities, search for single pulses, and classify the candidates as radio frequency interference (RFI) or potential real bursts (Methods).

In order to investigate the presence of X-ray bursts from SGR~1935$+$2154, we searched the \href{https://heasarc.gsfc.nasa.gov}{\textsc{heasarc}} archive for X-ray observations performed simultaneously with our radio observations. We found relevant overlap with our radio campaign at \rm{NICER}, \rm{Swift}, and \rm{Fermi} (see Methods for details). Finally, we considered the \href{http://enghxmt.ihep.ac.cn/dqjh/317.jhtml}{observing schedule} and \href{http://enghxmt.ihep.ac.cn/bfy/331.jhtml}{burst list}\cite{li_burstcatalogue_hmxt} from the \rm{Hard X-ray Modulation Telescope} (\cite[\rm{HXMT}][]{hxmt_2020}).

\begin{table*} 
\caption{\label{tab:coverage}Observational setup}
\begin{tabular}{lccccc}
\hline
\hline
Station$\mathrm{^{a}}$  & Band$\mathrm{^{b}}$  & Bandwidth [MHz]$\mathrm{^{c}}$ & SEFD [Jy]$\mathrm{^{d}}$ & Completeness [Jy$\,$ms]$\mathrm{^{e}}$ & Time observed [hrs]$\mathrm{^{f}}$ \\
\hline
Wb  & P            & 40               & 2100  & 78   & 102.6\\
Wb  & L$_{\rm Wb}$ & 100              & 420   & 10   & 278.8 \\
O8  & L$_{\rm O8}$ & 100, 175, 250 & 350   & 8, 6, 5 & 208.5 \\
Tr  & C        & 240      & 220            & 3       & 151.0 \\
O6  & X        & 500      & 785        & 8           & 22.4  \\
\hline
\multicolumn{5}{l}{Total telescope time/total time on source [hrs]$\mathrm{^{g}}$} & 763.3/522.7 \\
\hline

\multicolumn{6}{l}{$\mathrm{^{a}}$ Wb: Westerbork RT1, O8: Onsala 25m, Tr: Toru\'n, O6: Onsala 20m.} \\
\multicolumn{6}{l}{$\mathrm{^{b}}$ P: 314--377$\,$MHz; L$_{\rm Wb}$: 1260--1388$\,$MHz; L$_{\rm O8}$: varying ranges between 1227--1739$\,$MHz, see full details in Table \ref{tab:observations};} \\
\multicolumn{6}{l}{\hspace{0.5cm}C: 4550--4806$\,$MHz; X: 8080--8592$\,$MHz.} \\
\multicolumn{6}{l}{$\mathrm{^{c}}$ Effective bandwidth accounting for RFI and band edges.} \\
\multicolumn{6}{l}{$\mathrm{^{d}}$ From the \href{http://old.evlbi.org/user_guide/EVNstatus.txt}{EVN status page.}} \\
\multicolumn{6}{l}{$\mathrm{^{e}}$ Assuming a $7\sigma$ detection threshold} \\
\multicolumn{6}{l}{$\mathrm{^{f}}$ Please see Fig. \ref{fig:obssummary} and Table \ref{tab:observations} for exact time ranges of the observations.} \\
\multicolumn{6}{l}{$\mathrm{^{g}}$ Total time on source accounts for overlap between the participating stations.} \\
\end{tabular}
\end{table*}

\section{Results}\label{results}
We detected two bursts in the data from Wb (central observing frequency $\nu=1324.0\,$MHz, Table \ref{tab:coverage}) on 2020 May 24 at barycentric arrival times 22:19:19.67464 UT and 22:19:21.07058 UT
(B1 and B2, respectively, dispersion corrected to infinite frequency).
{\tt Heimdall} detected the bursts at a S/N of $81.9$ for B1 and $24.6$ for B2. FETCH (model A), in turn, reports a probability of 1.0 for both bursts to be of
astrophysical origin.

\begin{figure*} 
\resizebox{\hsize}{!}
        {\includegraphics{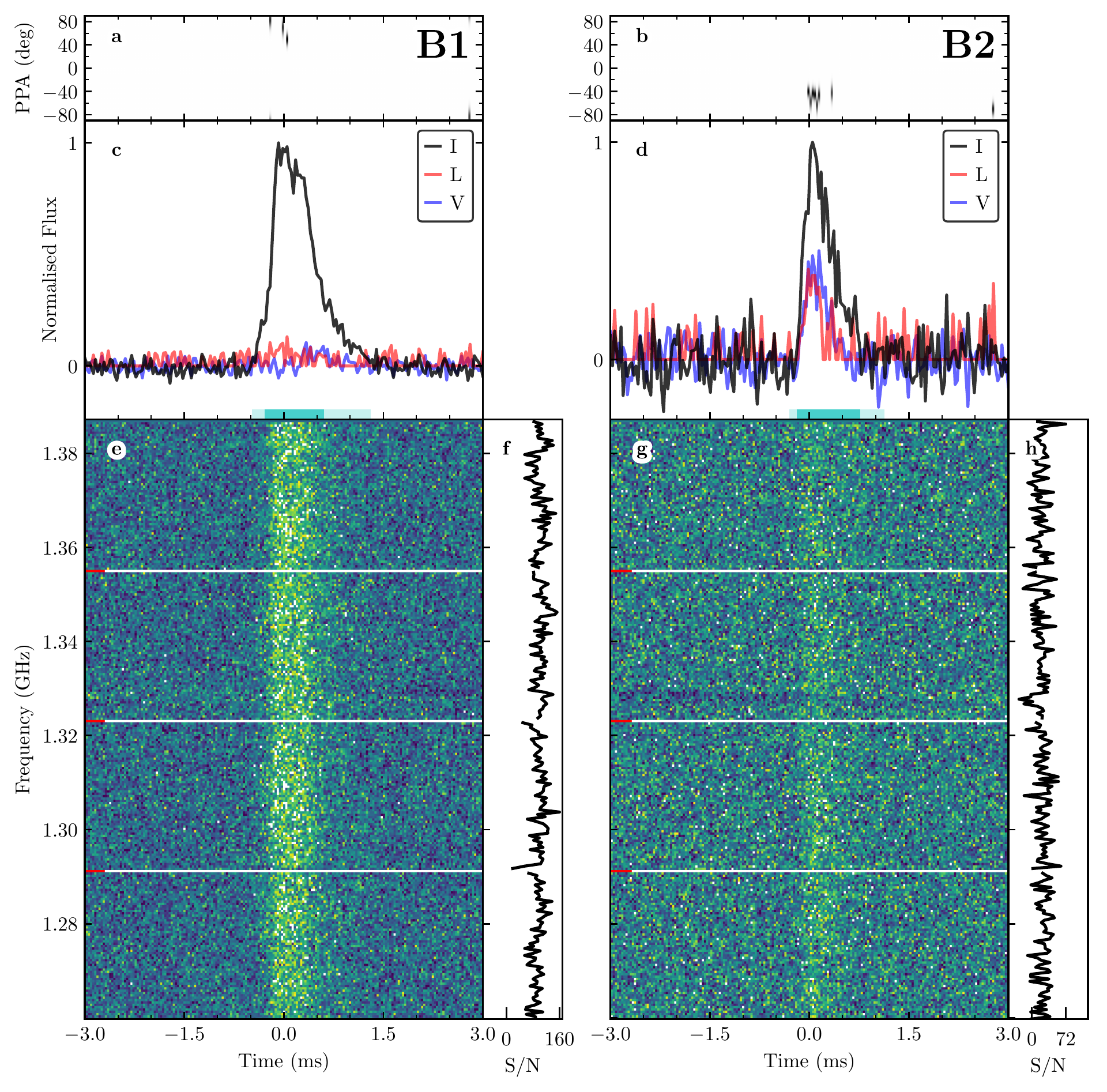}}
  \caption{Full polarisation profiles and dynamic spectra of the two bursts. B1 and B2 are displayed in the left and right column, respectively. The bursts are plotted with a time and frequency resolution of $32\,\mathrm{\upmu s}$ and $500\,\mathrm{kHz}$, respectively, and are coherently de-dispersed using a DM of $332.7206\,\mathrm{pc\,cm^{-3}}$. {\bf a} and {\bf b}: polarisation position angle. The greyscale represents the probability distribution of the PPA\cite{everett2001}, the darker shading representing higher polarised S/N. The PPA is masked below the linear S/N of $3$. {\bf c} and {\bf d}: band-averaged profiles. The dark cyan bars represent the full-width at half-maximum (FWHM; Table \ref{tab:burst_properties}) of the burst profile as determined with a Lorentzian fit to the autocorrelation function of the bursts in the time direction. The light cyan bars are $2$ and $1.5$ times the FWHM of B1 and B2, respectively. The cyan bars are placed such that they maximise the derived fluence. The total intensity burst profile is shown in black; the red and blue profiles represent the Faraday-corrected unbiased linear (see Equation \ref{eq:Lunbias}) and circular polarisation, respectively.  {\bf e} and {\bf g}: dynamic spectra. The white bands marked with red ticks in the dynamic spectra indicate frequency channels that have been masked due to subband edges. For visual purposes the limits of the colour map have been set to the $1^{\text{st}}$ and $99^{\text{th}}$ percentile of the dynamic spectrum. The dark bands in the 1.325--1.335$\,$GHz region are due to persistent RFI. {\bf f} and {\bf h}: time-scrunched, bandpass-corrected spectra computed as the sum of the dynamic spectrum under the light cyan bars in {\bf c} and {\bf d}. The displayed times are referenced to the arrival times listed in Table \ref{tab:burst_properties}.}
     \label{fig:burstsplot}
\end{figure*}

We subsequently create coherently dedispersed filterbanks with SFXC using\cite{chime_magnetar_2020_arxiv} $\mathrm{DM_{SGR}=332.7206\,pc\,cm^{-3}}$ (Methods). In Fig. \ref{fig:burstsplot} we show the resulting dynamic spectra and full-polarisation burst
profiles.
A coherently dedispersed filterbank with a time resolution of $8$ $\mu$s and a frequency resolution $500$ kHz is used to determine the arrival times, fluences, peak flux densities, spectral energy densities, intrinsic pulse widths, observed burst widths and scattering time scales. The dynamic spectra are summed over frequency to create a normalised time series. We fit a Lorentzian distribution to the autocorrelation function of the time series to determine the full-width at half-maximum (FWHM) of the burst profiles. The resulting observed burst widths are $866\pm43\,\mu$s for B1 and $961\pm48\,\mu$s for B2, and are shown using a dark cyan bar in Fig. \ref{fig:burstsplot}. The fluences of the bursts are determined by integrating over the light cyan bar shown in Fig. \ref{fig:burstsplot}, which have widths of $2$ and $1.5$ times the FWHM for B1 and B2, respectively. These factors were chosen such that the light cyan bars fully cover the entire burst envelope. The fluence and peak flux density are converted to physical units using the radiometer equation\cite{cordes_2003}, and the spectral energy density is determined assuming a distance to SGR~1935$+$2154 of $d=9.0\pm2.5\,\mathrm{kpc}$\cite{zhong_distanceSGR_2020arXiv200511109Z}. The burst properties are presented in Table \ref{tab:burst_properties}.
Given the system equivalent flux density (SEFD) and available bandwidth at
each station we estimate our burst searches to be complete to the  $7\sigma$-fluence
limits listed in Table \ref{tab:coverage}.

\begin{figure*} 
\resizebox{\hsize}{!}
        {\includegraphics{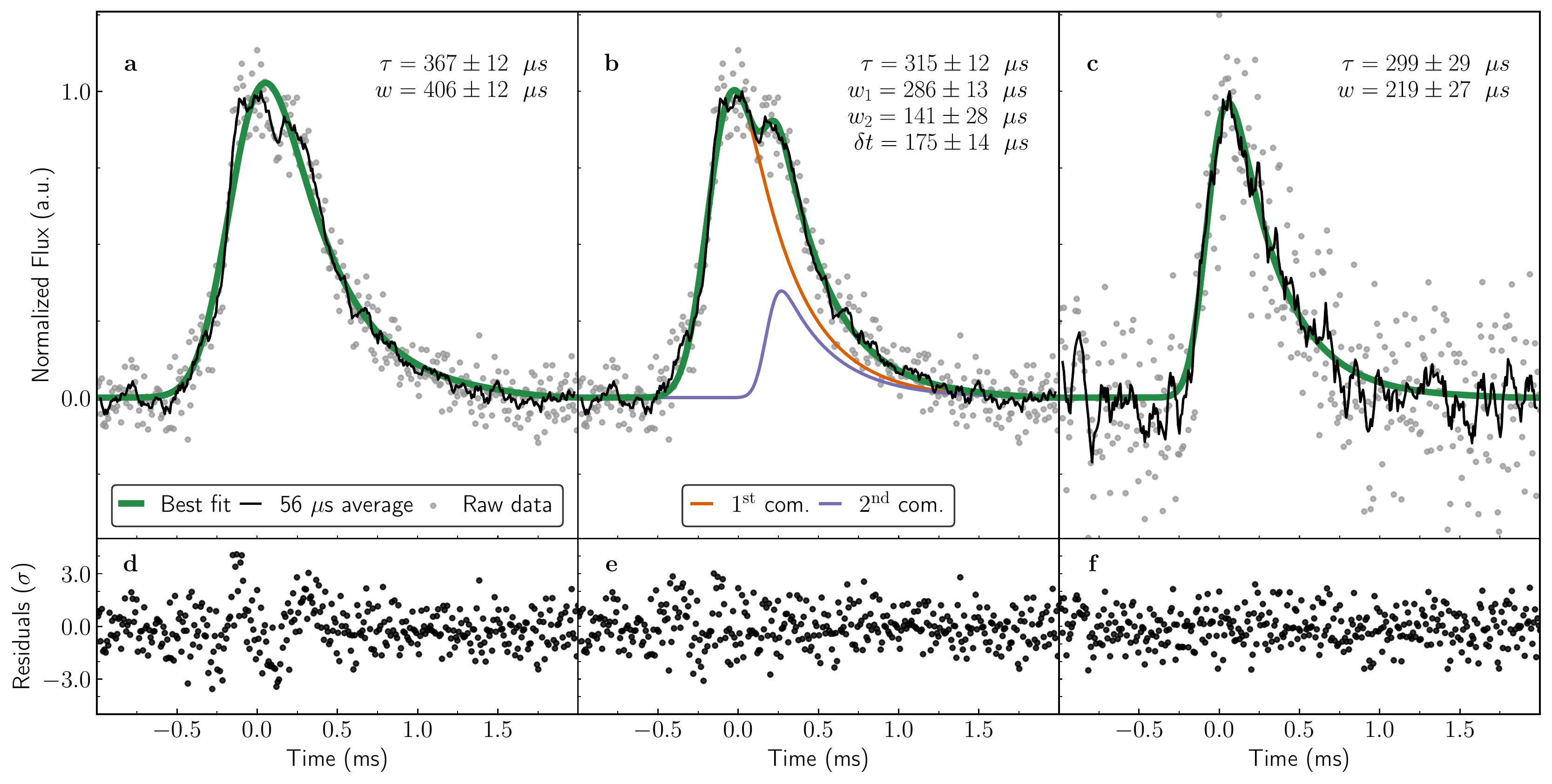}}
  \caption{Time series of the bursts. We fit a Gaussian distribution convolved with a one-sided exponential decay to both bursts. For B1 we attempted both a single-component fit (panel {\bf a}) and a two-component fit (panel {\bf b}) in which we keep the scattering time, $\tau$, the same for both components (orange and purple lines show the individual components). The bottom panels ({\bf d-f}) show the residuals for each fit. Gray dots display the raw data. The best fits are solid green lines and, for visual purposes, a $56$ $\mu$s running average has been plotted as a solid black line. Fitting results are displayed in each panel, where $\tau$ and $w$ denote the fitted scattering time scale and width of the Gaussian, respectively. Uncertainties denote the $1\sigma$ statistical errors of the fits. In panel {\bf b} we fit separate widths, $w_{1}$ and $w_{2}$, for each component and denote the delay between the peaks of the two Gaussians as $\delta t$. It is obvious from the residuals and the quality of the fits that a two-component model provides a much better fit to B1 (see text for details).
  }
     \label{fig:scatterfits}

\end{figure*}

\begin{table*} 
\caption{\label{tab:burst_properties}Burst properties}
\begin{tabular}{lcc}
\hline
\hline
    & B1   & B2 \\
\hline
Barycentric arrival time [MJD]$\mathrm{^{a}}$ & 58993.93008882 &  58993.93010498 \\
Dispersion measure [pc$\,$cm$^{-3}$]$\mathrm{^{b}}$   & $332.85\pm0.21$ & $332.94\pm0.21$\\
Fluence [Jy$\,$ms]$\mathrm{^{c,d}}$          & $112\pm22$         & $24\pm5$ \\
Peak flux density [Jy]$\mathrm{^{c}}$   & $170\pm34$ & $64\pm13$ \\
Spectral energy density [erg/Hz]$\mathrm{^{c,e}}$ & $(1.1^{+1.0}_{-0.6}) \times 10^{22}$ & $(2.3^{+2.2}_{-1.3}) \times 10^{21}$ \\
Intrinsic pulse width [$\mu$s]$\mathrm{^{f}}$         & $427\pm33\mathrm{^{g}}$      & $219\pm27$ \\
Observed burst width [$\mu$s]$\mathrm{^{h}}$         & $866\pm43$      & $961\pm48$ \\
Scattering time scale [$\mu$s] & $315\pm12$   & $299\pm29$ \\
Decorrelation bandwidth [kHz]  & $<500\,$Hz  & $<500\,$Hz \\
Linear polarisation L$_{\rm unbias}$/I [\%]$\mathrm{^{i}}$  & 8.3 $\pm$ 1  &   27.7 $\pm$ 2\\
Circular polarisation $\mid$V$\mid$/I [\%]$\mathrm{^{i}}$  &  7.7 $\pm$ 1  &  39.4 $\pm$ 3\\
\hline
\multicolumn{3}{l}{$\mathrm{^{a}}$ Time of arrival of the peak of the burst envelope at the Solar System}  \\
   \multicolumn{3}{l}{\hspace{0.3cm}Barycentre after correcting to infinite frequency using  } \\
   \multicolumn{3}{l}{\hspace{0.3cm}$\mathrm{DM=332.7206\,pc\,cm^{-3}}$.} \\
\multicolumn{3}{l}{$\mathrm{^{b}}$Determined using PSRCHIVE's {\tt pdmp}.} \\
\multicolumn{3}{l}{$\mathrm{^{c}}$Uncertainties are based on a 20\%-uncertainty in the system} \\
   \multicolumn{3}{l}{\hspace{0.3cm}temperature measurements.} \\
\multicolumn{3}{l}{$\mathrm{^{d}}$Integrated over the light cyan bar shown in Fig. \ref{fig:burstsplot}.} \\
\multicolumn{3}{l}{$\mathrm{^{e}}$Assuming a distance $d=9.0\pm2.5\,$kpc\cite{zhong_distanceSGR_2020arXiv200511109Z}.} \\
\multicolumn{3}{l}{$\mathrm{^{f}}$Defined as the FWHM of the Gaussian component before convolution.} \\
\multicolumn{3}{l}{$\mathrm{^{g}}$As per the sum of both widths from the 2-component fit in Fig. \ref{fig:scatterfits}.} \\
\multicolumn{3}{l}{$\mathrm{^{h}}$Defined as the FWHM of the Lorentzian distribution fitting the} \\
   \multicolumn{3}{l}{\hspace{0.3cm}autocorrelation function of the time series and using a } \\
   \multicolumn{3}{l}{\hspace{0.3cm}10\%-fractional error.} \\
\multicolumn{3}{l}{$\mathrm{^{i}}$Errors quoted are $1\sigma$ statistical errors that assume the errors on} \\
   \multicolumn{3}{l}{\hspace{0.3cm}the Stokes parameters are independent, and the errors are independent per time bin.} \\
   \multicolumn{3}{l}{\hspace{0.3cm}These uncertainties do not account for calibration errors and the effect of removing the baseline.}
\end{tabular}
\end{table*}

\subsection{Polarimetric properties of the bursts}\label{sec:pol}

We used full-polarisation data, with time and frequency resolution $32\,\mathrm{\upmu s}$ and $125\,\mathrm{kHz}$, respectively, to study the polarimetric properties of the bursts from SGR~1935$+$2154. In this analysis we did not perform a calibration scan to use for polarimetric calibration. Instead, we used our test pulsar observation, of PSR~J1935$+$1616, to determine the leakage correction (10\%) and the delay correction ($\sim$2$\,$ns) between the recorded right and left circular polarisation (Methods). We assume that there are no significant changes to the calibration required between the test pulsar scan and the detected bursts as the respective scans are less than 1$\,$hr apart.

We measure the rotation measure (RM) of B2 to be $\mathrm{RM_{B2}=107\pm18 \,rad\,m^{-2}}$ (Methods), consistent with the previous measurements\cite{chime_magnetar_2020_arxiv,fast_polarised_burst_zhang_2020ATel13699....1Z}.
For burst B1, however, we cannot measure the RM which we attribute to the double-component structure seen in B1 (Fig. \ref{fig:scatterfits}). The possibly two independent bursts overlap in time such that their polarisation properties are superimposed which, effectively, leads to a depolarised signal. We deem the depolarisation unlikely to arise from a significant change to the calibration solutions, since we find consistent results from PSR~J1935$+$1616 (before burst B1) and burst B2 ($1.4\,\mathrm{s}$ after B1).
Assuming that the RM has not changed significantly between the two bursts, i.e. the RM of burst B1 is consistent with B2, we use $\mathrm{RM_{B2}}$ to de-Faraday both B1 and B2.
In Fig. \ref{fig:burstsplot} we show the Faraday-corrected (Methods) polarisation profiles of both bursts, and the polarisation position angle, $\mathrm{PPA=0.5\arctan(U/Q)}$. For B2, the PPA is consistent with being flat across the burst profile, similar to what \citet{chime_magnetar_2020_arxiv} report. In Table \ref{tab:burst_properties}, we quote the linear and circular polarisation fractions for B1 and B2 determined by summing the polarisation profile and dividing by the sum of the Stokes I profile. The uncertainties quoted are 1$\sigma$ errors assuming the errors in the Stokes parameters are independent, and the error in each time bin is independent. The uncertainties quoted also do not encapsulate calibration uncertainties or the effect of removing the background from the data.

\subsection{Scattering and scintillation}
To determine the scattering times, a Gaussian profile convolved with an exponential decay, i.e. a thin screen model, is fit to each profile. As can be seen in the burst profiles of Fig. \ref{fig:burstsplot}, B1 exhibits a double-peaked structure. Therefore, we fit both a single and a double component burst to the profile of B1. For the double component fit, the decay time was fixed to be the same for both components. We find a reduced chi-square value $\chi^2_{\nu} = 1.6$ for the single component fit and $\chi^2_{\nu} = 1.2$ for the double component fit. Furthermore, the difference in the $\chi^2$ value, $\Delta \chi^{2}$ is $136$ for three additional degrees of freedom, which indicates that the double component fit is a $> 11 \sigma$ improvement over the single component fit. We conclude that B1 is consistent with exhibiting a double component temporal structure. For B2 we find $\chi^2_{\nu} = 1.0$. The double component fit for B1 and the single component fit for B2 result in scattering times $\tau_{\mathrm{B1}}=315\pm12\,\mu$s and $\tau_{\mathrm{B2}}=299\pm29\,\mu$s.
The weighted average is $\bar{\tau}=313\pm31\,\mu$s at $1324\,$MHz, where we added the uncertainties in quadrature. Within the model of a thin scattering screen, where the scattering time scale $\tau$ and the scintillation bandwidth $\mathrm{\nu_{scint}}$ are related via $2\pi\tau\nu_{\mathrm{scint}}=1$, our $\bar{\tau}$ implies a scintillation bandwidth of about 500$\,$Hz. An autocorrelation analysis of coherently dedispersed data with a frequency resolution of 488$\,$Hz yields no scintillation bandwidth larger than the width of one channel. Producing a filterbank with even higher frequency resolution would require a time resolution of $>4\,$ms, and would reduce the S/N of any apparent scintillation because this timescale is significantly longer than the burst duration.

\subsection{Burst rates}
\begin{figure} 
    \centering
    \includegraphics[width=\linewidth]{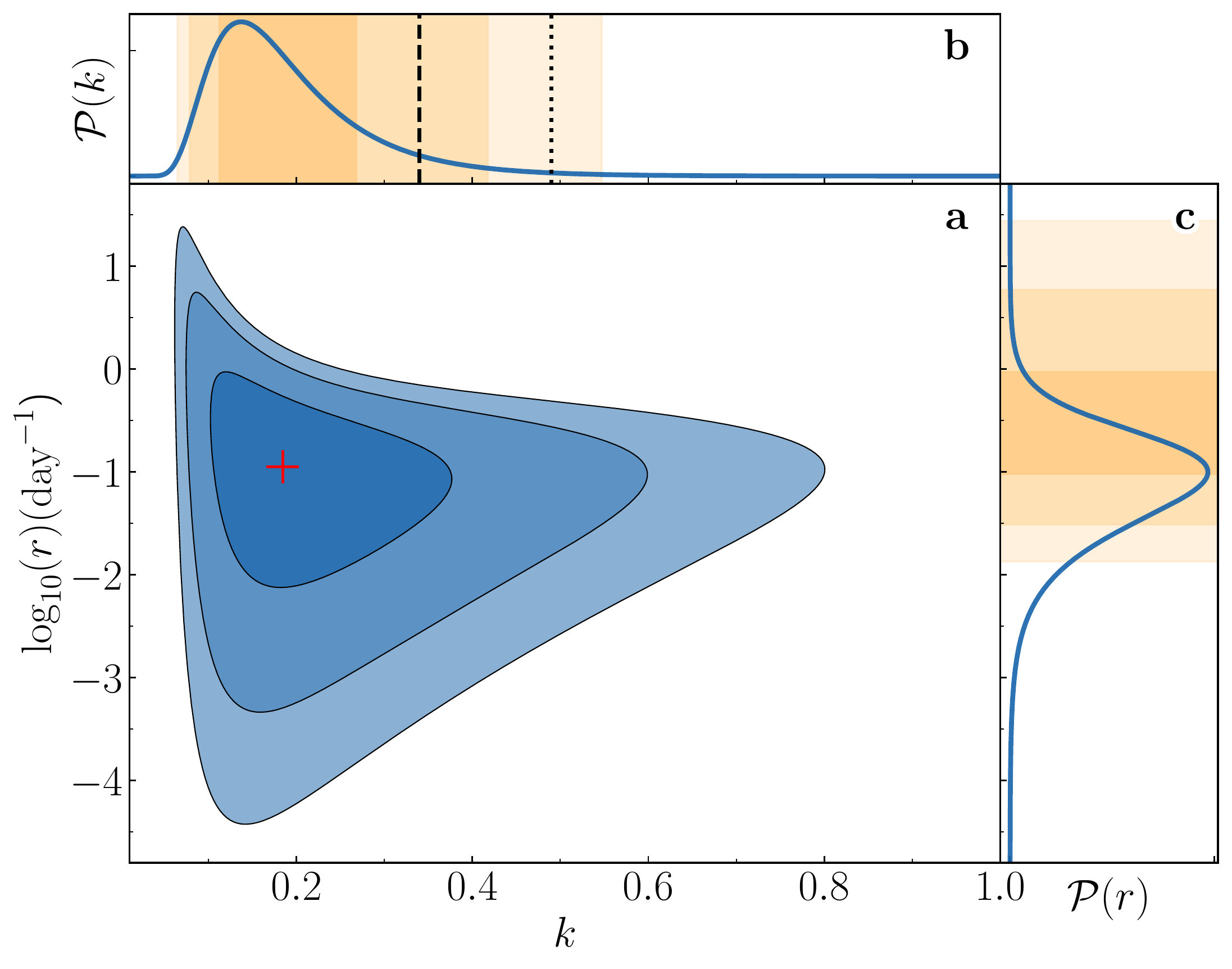}
    \caption{Posterior distribution of $k$ and $r$ parameters of the Weibull distribution. Panel {\bf a} shows the full 2D-distribution with the red cross corresponding to the point of maximum probability density, and the shaded blue contours represent the 68\%, 95\%, and 99\% confidence regions surrounding the maximum. The marginal distributions are shown in panels {\bf b} (for $k$) and {\bf c} (for $r$); the orange shading represents the 68\%, 95\%, and 99\% confidence intervals. The black dashed and black dotted lines indicate the best fit values of $k$ determined for FRB~121102\cite{Oppermann_2018,Oostrum_2020}. Note that the distribution of $r$ looks symmetric due to the logarithmic scale, but it is actually skewed towards larger values.}
    \label{fig:weibull}
\end{figure}
The time span between bursts B1 and B2 is only $\sim$1.4$\,$s, which is very short compared to the roughly 421$\,$hrs total duration of non-overlapping observations taken at L-band. Therefore, we assume a Weibull distribution\cite{Oppermann_2018} to estimate the burst rate $r$ and the shape parameter $k$ valid in this frequency band (Methods).  The most likely values of $k$ and $r$ taken jointly are $k=0.18$ and $r=0.11\,\mathrm{day^{-1}}$ (Fig. \ref{fig:weibull}). Moreover, the 68\% confidence interval for $k$ is 0.11--0.27, while the 68\% confidence interval for $r$ is 0.10--0.93$\,\mathrm{day^{-1}}$. Thus the data do not support a Poissonian model (for which $k=1$), and there is evidence for clustering. Interestingly, the 95\% confidence interval for $k$ (0.08--0.42)  is consistent with the 2$\sigma$-region for $k$ derived for FRB~121102\cite{Oppermann_2018,Oostrum_2020} (Fig. \ref{fig:weibull}). This is an intriguing similarity between repeating extragalactic FRBs and SGR~1935$+$2154, although we cannot draw inferences about the exact mechanism itself.

\subsection{X-ray bursts during the radio campaign\label{subsec:res:xray}}

The pointed \rm{Swift} and \rm{NICER} observations did not reveal any X-ray bursts from SGR~1935$+$2154. While the source was in the field of view of \rm{Fermi}/GBM during the two radio bursts on 2020 May 24, no simultaneous X-ray bursts were detected. \rm{HXMT} was not observing SGR~1935$+$2154 during the radio bursts \citep{li_burstcatalogue_hmxt}, while the source was not in the \rm{Swift}/BAT field of view at that time.

On the other hand, several X-ray bursts were observed overlapping with our radio monitoring, without an associated radio burst detection. No radio bursts\cite{kirsten_2020ATel13735} were seen during the X-ray burst detected with several X-ray instruments\cite{may10_burst_gcn} on 2020 May 10. Similarly, no radio burst was observed when \rm{Fermi} triggered on a SGR~1935$+$2154 burst on 2020 May 20 (event bn200520908). We fit the spectrum of this burst with a double blackbody model (\textsc{bbody}+\textsc{bbody} in \textsc{xspec}), adding a cross-correlation multiplication constant between the spectra from detectors n3, n6, and n7. We measure temperatures of $kT_{\rm BB,1} = 5.2 \pm 0.4\,$keV and $kT_{\rm BB,2} = 16.7^{+6.7}_{-3.8}\,$keV for a fit with $\chi^2_{\nu} = 137.8/129 = 1.07$. We measure a 8--200$\,$keV fluence of $(3.6\pm0.3)\times10^{-7}\,\mathrm{erg\,cm^{-2}}$.

Comparing the \rm{HXMT} burst list with the radio campaign, we find 59 X-ray bursts overlapping the radio observations (see Table \ref{tab:hxmtbursts}). None of these are accompanied by a radio burst. At the time of writing, no information beyond fluence and $T_{90}$ values are reported for these bursts \citep{li_burstcatalogue_hmxt}. The brightest of these 59 overlapping X-ray bursts had a fluence of $2.01\times10^{-6}\,\mathrm{erg\,cm^{-2}}$, significantly brighter than the \rm{Fermi} burst discussed above. The faintest of these X-ray bursts, on the other hand, had a reported fluence of $8.64\times10^{-12}\,\mathrm{erg\,cm^{-2}}$.

\section{Discussion}\label{discussion} 

\subsection{Completeness and burst energy distribution}
\begin{figure} 
    \centering
    \includegraphics[width=\linewidth]{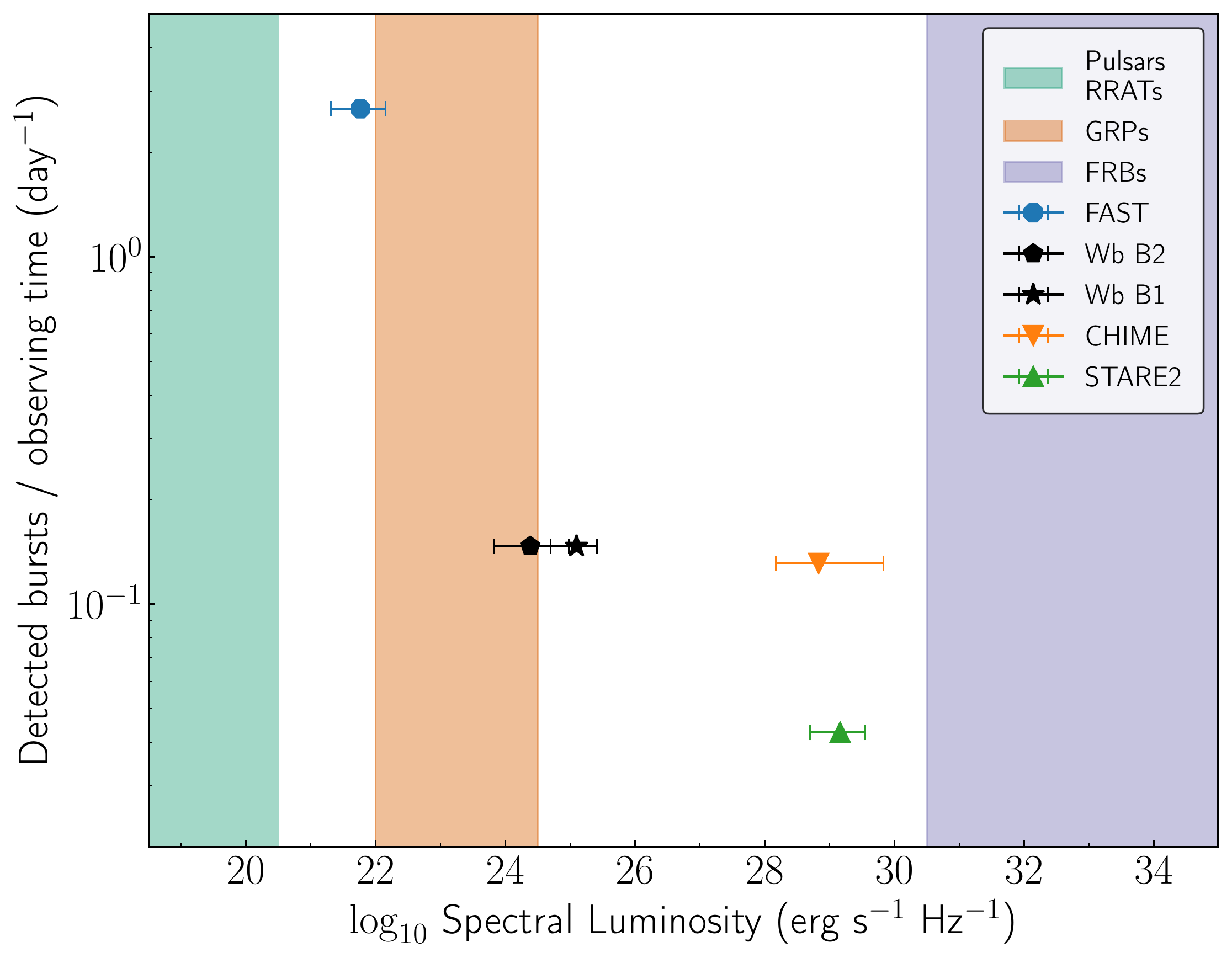}
    \caption{Burst occurrence per observing time and associated spectral luminosity. We display estimates for FAST\cite{fast_polarised_burst_zhang_2020ATel13699....1Z} (blue octagon), CHIME/FRB\cite{chime_magnetar_2020_arxiv} (orange downward triangle), STARE2\cite{bochenek_2020_arxiv} (green upward triangle) and this work (black star for B1, black pentagon for B2). The shaded regions indicate typical spectral luminosities for pulsars and rapidly rotating astrophysical transients (green), giant radio pulses (light brown), and FRBs (purple)\cite{bochenek_2020_arxiv}. It is obvious that the bursts reported for SGR~1935$+$2154 cannot be assigned to any of the three categories alone and that their detection rates are confined to a region spanning less than 3 orders of magnitude while the burst energies span close to eight orders of magnitude. (Methods)}
    \label{fig:spectral_luminosity}
\end{figure}

The rate and shape parameter determined above are valid for bursts brighter than our detection threshold of 8--10$\,\mathrm{Jy\,ms}$ at L-band (Table \ref{tab:coverage}). It is possible that we have missed bursts of lower fluence, e.g. bursts like the one reported by \citet{fast_polarised_burst_zhang_2020ATel13699....1Z}. On the other hand, the two bursts that we see within 522.7$\,$hrs on source are well above our detection threshold. In combination with the first-known burst\cite{chime_magnetar_2020_arxiv, bochenek_2020_arxiv}, which is also the only one detected within hundreds of hours of observations with CHIME/FRB and STARE2, this is indicative of an almost flat cumulative distribution function of burst energies (Fig. \ref{fig:spectral_luminosity}). Assuming that a single emission mechanism is responsible for all reported radio bursts from SGR~1935$+$2154, it has to be of such type that the burst rate is close to independent of the amount of energy emitted across more than seven orders of magnitude. Alternatively, different parts of the emission cone might cross our line of sight in case the beaming direction changes significantly over time.

We note that there also exists an upper detection threshold, which we estimate to be of order 10$\mathrm{\,kJy\,ms}$ for our system. Any signal above this fluence could lead to non-linearities in the receiver system causing us to miss such bursts. However, during the time range of our observations neither CHIME/FRB nor STARE2 reported further bursts as bright as FRB~200428. In addition, there is no S/N-limit above which signals are masked as RFI in our analysis. Thus, it appears unlikely that we have missed any extremely bright bursts during our observations.  

\subsection{Polarimetry}
\citet{fast_polarised_burst_zhang_2020ATel13699....1Z} presented the detection of a low-fluence, highly linearly polarised burst from SGR~1935$+$2154 with no circular polarisation detected. This is in contrast to the polarisation properties of the two bursts presented in this work. Our B2 is significantly less strongly polarised compared to the L/I $\sim 100\%$ of the FAST-detected burst, and B1 exhibits no significant polarisation ($<10\%$) at all.
Radio magnetars show a wide range of polarisation properties\cite{kramer_2007,camilo_2008,levin_2012}; it is possible, perhaps unsurprising, that B1 and B2 are intrinsically not $\sim 100\%$ polarised. However, we find evidence for scattering in the burst profiles of both B1 and B2 which could lead to partial depolarisation\cite{camilo_2008,levin_2012}.
Alternatively, the low linear polarisation observed in burst B1 can be caused by the superposition of the polarisation properties of the two closely spaced sub-bursts (Fig. \ref{fig:scatterfits}).

A diverse range of polarisation properties are also observed for FRBs, with linear polarisation fractions ranging from $\sim 0$ to $100\%$\cite{petroff_2015,masui_2015,ravi_2016,michilli_2018_magnetoionic}.

\subsection{Simultaneity of X-ray and radio bursts}
During the CHIME/FRB and STARE2 radio burst, with an estimated fluence of $1.5\pm0.3\,\mathrm{MJy\,ms}$ at 1378$\,$MHz\cite{bochenek_2020_arxiv}, an X-ray burst with a fluence in the range of $\sim$6.1--9.7$\mathrm{\times10^{-7}\,erg\,cm^{-2}}$ was detected by \rm{INTEGRAL}, \rm{Konus-Wind}, and \rm{HXMT} (in different energy ranges between 1 and 500$\,$keV\cite{Mereghetti_integral,ridnaia_2020_arxiv,li_insight_hxmt_simultaneous_CHIME_STARE2_Xray_burst_2020_arxiv}; note that \rm{AGILE} also detected the burst but has not yet reported a fluence measurement). Our brightest burst seen on 2020 May 24, B1, had a fluence four orders of magnitude weaker than the burst seen by STARE2. Assuming a similar ratio\cite{Mereghetti_integral} between radio and X-ray fluence during both bursts ($\sim$$10^{-5}$), we would expect a fluence of the order of $10^{-10}\,\mathrm{erg\,cm^{-2}}$ in X-rays. As this value is orders of magnitude lower than typical detection thresholds for \rm{Fermi} (of the order of $10^{-7}\,\mathrm{erg\,cm^{-2}}$ for $\sim$1$\,$s bursts\cite{fermi_UL1,fermi_UL2}), it is not surprising that \rm{Fermi} detects no X-ray bursts during the radio bursts.

Conversely, another three bright X-ray bursts coincident with our campaign were reported and a further 59 overlapping bursts are listed in Table \ref{tab:hxmtbursts}. We found no radio counterparts to any of these bursts in our radio observations\cite{kirsten_2020ATel13735}, which allows us to place upper limits on the radio fluences --- as listed in Table \ref{tab:coverage}. \citet{lin_fast_2020_arxiv} also report a non-detection of pulsed radio emission in an observing campaign with FAST, during which 29 high energy bursts were reported by the Fermi Gamma Ray Burst monitor. Therefore it seems that the majority of X-ray/Gamma ray bursts are not associated with pulsed radio emission. The parameters and fluences that we measure for the X-ray bursts discussed in Section \ref{subsec:res:xray} are consistent with typical values observed for SGR~1935$+$2154 \citep{linlin2020}, fitting with the idea that radio bursts are instead associated with atypical, harder X-ray bursts \citep{younes_NICER_GBM}.

\subsection{Implications for magnetars and FRBs}

To date, five Galactic magnetars, all of which are considered `transient magnetars', have shown pulsed radio emission \citep{2017ARA&A..55..261K,Lower2020}.  This emission is transient, lasting weeks to months, and associated with an X-ray outburst.
In comparison to the radio-pulsing magnetars, SGR~1935$+$2154 produces {\it much} more sporadic bursts, and suggests that high-cadence monitoring of other Galactic magnetars
might also discover radio bursts associated with X-ray burst storms.  Along with SGR~1935$+$2154, the discovery of two bright, sporadic bursts from the radio-emitting magnetar J1550$-$5418 \citep{Burgay2018} strengthens the idea that this may not be uncommon.

The 1.396-s separation between bursts B1 and B2 corresponds to 0.43 of SGR~1935$+$2154's 3.245-s rotational period. Currently it is impossible to assign rotational phases to our and all other detected radio and X-ray bursts from SGR~1935$+$2154 due to the lack of a phase-coherent rotational ephemeris.
This is important though for understanding the burst emission mechanism. It might contribute to understanding the apparent lack of burst arrival time periodicity from repeating FRBs, which could in principle be attributed to bursts occurring at a wide and varying range of rotational phases \citep{2016Natur.531..202S}, i.e. from varying emission sites --- as opposed to being from a relatively stable location of origin, as is the case in rotation-powered radio pulsars.
Our SGR~1935$+$2154 results suggest that its bursts can occur at a wide range of rotational phases, but with only two bursts we can not rule out a more stable pulse-interpulse configuration.

The four reported radio bursts from SGR~1935$+$2154 span more than seven orders of magnitude in observed fluence.  While beaming of the radio emission certainly must affect the observed fluences at some level, this nonetheless demonstrates that SGR~1935$+$2154's radio burst emission spans the typical luminosities seen from rotation-powered radio pulsars up to the closest-known extragalactic FRBs (see Fig. \ref{fig:spectral_luminosity}).  It is unclear whether the four known SGR~1935$+$2154 bursts were produced by the exact same type of physical process.
Neutron stars are known to produce radio bursts of various types (polar-cap pulsar emission, giant pulses, radio magnetar emission).  Perhaps the observational differences between the bursts from repeating and (apparently) non-repeating sources are also a reflection of this diversity of emission mechanisms seen from neutron stars.

Observationally, one can pose the question: are low-luminosity radio bursts, that can only be detected from a Galactic source, also `FRBs'?  The repeater FRB~121102 has been observed to produce radio bursts with fluences spanning three orders-of-magnitude; for FRBs in general, the detection of lower/higher fluences is limited by telescope sensitivity and available observing time, respectively.

Overall, SGR~1935$+$2154 makes a compelling case that there is a link between (at least some) FRBs and magnetars.  However, important observational differences remain.  For instance, some repeating FRBs have shown periodicity in their activity level\cite{2020Natur.582..351C} on timescales of weeks to months --- suggesting that the source may be in a binary system, extremely slowly rotating, or rapidly precessing\cite{Lyutikov2020_binary,Beniamini2020,levin_FRBprecession_2020ApJ...895L..30L}.  SGR~1935$+$2154 is not known to be in a binary, and there are not yet enough detected radio bursts to look for a periodicity in its radio burst activity.  Using 174 X-ray bursts detected from 2014--2020, \citet{Grossan2020} claim periodic windowed activity with a period of 232$\,$days and a fractional activity window of 56\%.  Continued radio monitoring of SGR~1935$+$2154 can help verify this claim.

So SGR~1935$+$2154 is not a flawless analogue of the extragalactic FRB population.  Nonetheless, magnetars can plausibly explain the diverse phenomena observed from FRBs.  Perhaps the distant, periodically active FRB sources are brighter and more active because they are significantly younger than SGR~1935$+$2154 and because their magnetospheres are perturbed by the ionised wind of a nearby companion.  Similarly, perhaps non-repeating FRBs are older, non-interacting, and thus less active.  Detailed characterisation of FRB local environments is critical to investigating these possibilities.

\section{Methods} 
\subsection{Observations}\label{obs}
\subsubsection{Radio observations}\label{sec:observations:radio}

\begin{figure*} 
\resizebox{\hsize}{!}
        {\includegraphics{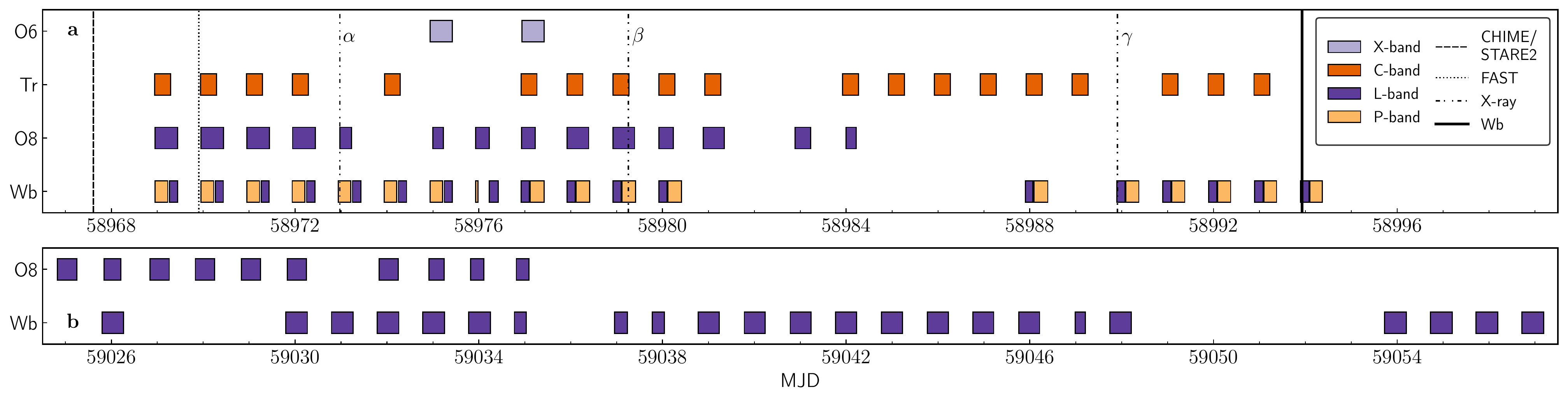}}
  \caption{Overview of the observations of SGR~1935$+$2154  during this campaign. Panels {\bf a} and {\bf b} both span 33$\,$days, with observations colour-coded by observing frequency. Note the gap of 25 days between {\bf a} and {\bf b}. No observations were conducted during that time period. Vertical lines indicate the times of reported bursts. Solid line: events found in our campaign; long-dashed: CHIME and STARE2 detections \citep{scholz_CHIME_2020ATel13681, bochenek_2020ATel13684_stare2}; dotted: detection by FAST  \citep{fast_polarised_burst_zhang_2020ATel13699....1Z}; dash-dotted: X-ray bursts as reported by $\alpha$) \citet{ursi_may3_burst_2020GCN.27687....1U}, $\beta$) \citet{may10_burst_gcn} and \citet{agile_may_10_xray_burst_2020GCN.27727....1V}, $\gamma$) a Fermi/GBM trigger on 2020 May 20 at 21:47:07.548 UT. During X-ray events $\beta$) and $\gamma$) no radio counterparts were found in any of our data, which allows us to place upper limits on the fluences --- as listed in Table \ref{tab:coverage}. Unfortunately we can draw no conclusions from our data coincident with event $\alpha$) because Wb was in a recording gap and O8 was affected by strong RFI.}

     \label{fig:obssummary}
\end{figure*}

Since the announcement of FRB~200428\cite{scholz_CHIME_2020ATel13681,bochenek_2020ATel13684_stare2}, we observed SGR~1935$+$2154 daily for up to almost
12 hours,
between 2020 April 29 UT 22:45 (MJD$\,$58968.94791) and 2020 May 25 UT 09:00 (MJD$\,$58994.37500).  After detecting bursts, we resumed the campaign with a similar cadence between 2020 June 24 UT 19:30 (MJD$\,$59024.81250) and 2020 July 27 UT 04:30 (MJD$\,$59057.18750).  See Fig. \ref{fig:obssummary}, Table
\ref{tab:coverage} and Table \ref{tab:observations} for a summary of the observing campaign. The telescopes involved were the 25-m single dish RT1
at Westerbork in the Netherlands (Wb, P- and L-band), the 25-m and 20-m telescopes at Onsala Space Observatory in Sweden (O8, O6; L- and X-band) and the 32-m dish in Toru\'n, Poland (Tr, C-band). All stations operated independently as single dishes,
recording 2-bit baseband data (circular polarisations) in VLBI Data Interchange Format
\citep[VDIF,][]{2010ivs..conf..192W} with the local Digital Base Band Converters (DBBC2 or
DBBC3 systems). In total, we observed for 763$\,$hrs, which reduces to 522.7$\,$hrs on source when accounting for overlap between the stations.

{\bf Westerbork RT1}: Wb observed in two different frequency ranges, covering 313.49--377.49$\,$MHz (P-band) split
into eight 8$\,$MHz wide subbands during
part of each run. The other part of a run covered 1260--1388$\,$MHz (L-band) split
at first into four 32$\,$MHz wide bands (29 April--19 May). This was changed to eight 16$\,$MHz wide
bands for easier processing as of 20 May. We recorded 3-minute scans with a 1-minute gap in between scans
during the first seven runs (29 April--06 May); for the remaining observations this was changed to
10-minute recordings and 20-second gaps. At the beginning of both the P-band observations and the L-band
observations we observed either pulsar PSR~J1921$+$2153 or the pulsar PSR~J1935$+$1616 as test sources to verify the system.

{\bf Onsala}: The Onsala 25-m dish (O8) observed at L-band with varying frequency ranges and bandwidths over
14 nights. We
recorded the entire available bandwidth of 512$\,$MHz between 1222--1739$\,$MHz during the first three observations
(29 April--02 May). Owing to the large fraction of radio frequency interference (RFI, $\sim$50\%) in the band
we subsequently tested setups with 256$\,$MHz of continuous bandwidth placed within the above range (02--09 May).
Eventually, we settled for a 128$\,$MHz wide band split into eight 16$\,$MHz wide bands between 1360--1488$\,$MHz
for the rest of the campaign (Table \ref{tab:observations}). We observed either PSR~J0358$+$5413 or PSR~J1935$+$1616 as test sources
towards the beginning of the observations.
For two runs (06--08 May) the Onsala 20-m telescope (O6) joined the observations covering the frequency range
8080--8592$\,$MHz (X-band), split into sixteen 32$\,$MHz wide subbands.
Both stations O8 and O6 observed for five to twelve hours during each run, recording
15-minute scans with a 12-second gap in between scans.

{\bf Toru\'n}: The 32-m dish at Toru\'n (Tr) observed at C-band for about 8$\,$hrs during a total of 19 nights.
We recorded the entire 256$\,$MHz of bandwidth covering the frequency range of 4550--4806$\,$MHz, split into eight
32$\,$MHz wide subbands. We performed 5-minute scans on the tests pulsars at the beginning and
the end of each observing run. During the first six nights (29 April--5 May) we scheduled a main 15-minute observing loop
that consisted of 880$\,$s of recording on SGR~1935$+$2154 and 20-second gaps dedicated to gain correction. For these first runs we observed PSR~J1935$+$1616 and PSR~J2022$+$2854 as the test sources. Thereafter we increased the gaps by 10$\,$s but the length of the observing loop was left unchanged. Also, from 7 May onward only PSR~J2022$+$2854 was observed for the system performance checking. We also observed during the night of 3 May 2020 for which
\citet{li_may3_burst_2020GCN.27688....1L} reported a bright X-ray burst, but due to a wrong setup the antenna was off source, hence all data were discarded.

\subsubsection{X-ray data}
Publicly available pointed observations were taken by the Neutron star Interior Composition ExploreR (\cite[\rm{NICER}][]{Gendreau2016_NICER}) and the Neil Gehrel's Swift Observatory (\cite[\rm{Swift}][]{Gehrels2004_Swift}), observing SGR~1935$+$2154 seven (ObsIDs 3020560107/8/25/33/37/40/42) and ten (ObsIDs 00033349049/50/56/58/60-63/66/76) times during the radio campaign, respectively. In addition, the target was in the field of view of the monitoring instruments aboard \rm{Swift} (the Burst Alert Telescope or BAT) and \rm{Fermi} (\cite[the Gamma-ray Burst Monitor or GBM][]{meegan_GBM_2009ApJ...702..791M}) the majority of the time. \rm{Swift}/BAT did not report any burst triggers during the radio observations. \rm{Fermi}/GBM records in Time-tagged Event (TTE) mode with a high,  $2\,\mu$s time resolution. Therefore, we focused on the \rm{Fermi}/GBM data at times of particular interest in the radio campaign.

\subsection{Data reduction and analysis}\label{analysis}
\subsubsection{Radio observations}
The baseband data from each participating station was transferred via the internet to Onsala Space Observatory (OSO)
where we searched the data from each station separately with a pipeline
that was developed to
search for FRBs in baseband recordings. We performed the following steps on a per station basis for each recorded scan:
\begin{enumerate}
    \item Create separate (baseband) files for each subband;
    \item Channelise each subband and form Stokes I;
    \item Splice all subbands together into one filterbank;
    \item Dedisperse the filterbanks and search for bursts;
    \item Classify and inspect burst candidates;
    \item Create coherently dedispersed filterbanks for the best candidates and verify.
\end{enumerate}
In the current recording setup the electric voltages are sampled as 2-bit real numbers. At each station each scan (i.e. each integration lasting 3--15$\,$minutes, rf. Section \ref{sec:observations:radio}) is recorded in a single
VDIF-file that contains both polarisations of all $N$ subbands (e.g. $N=8$ for Toru\'n which recorded eight 32$\,$MHz wide subbands to capture 256$\,$MHz of bandwidth). The software package
that we use to channelise
the baseband data and create total intensities (\cite[{\tt digifil}
from DSPSR][]{vanStraten_digifil_2011}) can
currently only unpack VDIF files that contain two polarisations of one single subband. Therefore, prior
to creating 8-bit filterbanks with {\tt digifil} we use  \href{https://github.com/jive-vlbi/jive5ab}{{\tt jive5ab}}  to split each scan into $N$
separate files that contain both circular
polarisations. Each subband is processed separately (but simultaneously) and the resulting
filterbank files are combined in one single file that contains the entire observed frequency range with
the utility {\tt splice} from SIGPROC\cite{lorimer_sigproc_2011ascl.soft07016L}. The time
resolution of the filterbanks at L-, C-, and X-band is $64\,\mu$s while the frequency resolution is 125$\,$kHz, 250$\,$kHz and 2$\,$MHz, respectively. Given the dispersion measure (DM) of SGR~1935$+$2154
($\mathrm{DM_{SGR}=332.7206\pm0.0009\,pc\,cm^{-3}}$\cite{chime_magnetar_2020_arxiv}),
this implies a maximal intra-channel time smearing of $<190\,\mu$s in our lowest channel at L-band
(1227$\,$MHz). The filterbanks created from the P-band data have a much finer channelisation (7.8$\,$kHz) to
limit residual intra-channel time smearing to $\sim$700$\,\mu$s at the lower end of the band. Time resolution is
accordingly lower (1$\,$ms) than in the other bands.

We manually inspect subsections of the data from each station to identify frequency ranges that are continuously affected by RFI. Based on this analysis, we create channel masks for flagging that are passed on to all subsequent steps of the burst search pipeline.

We search the filterbanks for bursts with \href{https://sourceforge.net/projects/heimdall-astro/}{{\tt Heimdall}}
as the dedispersion and burst finder engine. Since the dedispersion is known {\it a priori} we do not perform a
full search in DM-space but instead limit the search range to $\mathrm{DM_{SGR}\pm50\,pc\,cm^{-3}}$. The
candidates found by {\tt Heimdall} above a signal-to-noise (S/N) threshold of seven are then classified either as RFI or
potential candidates by FETCH (model A)\citep{agarwal19_fetch_arxiv}. We chose this particular S/N-threshold since while testing the pipeline a lower threshold led to an extensive number of false positives. This is easily explained by the fact that FETCH was trained and tested on data with S/N$\,\geq8$, i.e. the classifier employed by FETCH is potentially less reliable for low S/N-candidates. We inspect the candidates by eye and, as a
final step, we use the software correlator SFXC \citep{keimpema_SFXC_2015} to create coherently dedispersed
filterbanks around the times of the most convincing candidates, for final verification.

As mentioned above, we observed well-known pulsars in each observing run to verify the integrity of our data and
the reliability of our processing pipeline. To that end, we perform the steps described above also on the pulsar scans. In
addition, we fold the filterbank files that contain a scan of a pulsar
with DSPSR's {\tt dspsr} and inspect the folded profiles.
The respective pulsars were detected each time with the exception of PSR~J1921$+$2153 observed with station Wb at L-band. At
this frequency the pulsar was detected only about half the time, which we attribute to diffractive scintillation from the Galactic interstellar medium.
The test pulsar PSR~J1935$+$1616 is bright enough to detect several individual
pulses with our pipeline almost each time it is observed.

\subsubsection{Dispersion measure optimisation}
To optimise the DM we run the PSRCHIVE \citep{vanStraten_psrchive_2011} tool {\tt pdmp} on the filterbank data of each burst separately, which yields $\mathrm{DM_{B1}=332.85\pm0.21\,pc\,cm^{-3}}$ and $\mathrm{DM_{B2}=332.94\pm0.21\,pc\,cm^{-3}}$ for B1 and B2, respectively.
These values are consistent with DM$_{\mathrm{SGR}}$ as measured by \citet{chime_magnetar_2020_arxiv} ($\mathrm{DM_{SGR}=332.7206\pm 0.0009)\,pc\,cm^{-3}}$), albeit marginally higher. We attribute
the higher DM to the optimisation algorithm employed by {\tt pdmp} which
essentially maximises the S/N of the burst by modifying the DM. Given the
scattering tails of the bursts, this can lead to a peak in S/N at a DM higher than the true value.  We do not attempt to determine an optimal DM based on higher-time-resolution baseband data because the burst width is dominated by scattering.  Furthermore, we consider DM$_{\rm SGR}$ to likely be more accurate, because of the larger fractional bandwidth of those observations.

\subsubsection{Burst statistics}
If a stochastic process can be described as a Poisson point process with a constant rate parameter $r$, then the random variable describing the wait times $\delta$ between events generated by the process will follow an exponential distribution,
\begin{equation}
    f(\delta|r) = re^{-r\delta}.
\end{equation}
On the contrary, repeating FRBs are known to show clustering in their burst patterns, and therefore cannot be described with a Poissonian model. As described in \citet{Oppermann_2018}, a possible generalisation of the wait time distribution is given by the Weibull distribution,
\begin{equation}
    f(\delta|k, r) = \frac{k}{\delta}\big{(}\delta{r}\Gamma(1 + k^{-1})\big{)}^{k}e^{-\left(\delta{r}\Gamma(1 + k^{-1})\right)^{k}}
\end{equation}
with shape parameter $k$ and rate parameter $r$, which reduces to an exponential distribution if $k=1$. Here $\Gamma$ is the Gamma function. The posterior distribution of $k$ and $r$ can therefore be used to test whether the data supports a Poissonian model, because Poissonian data should necessarily produce a posterior distribution consistent with $k=1$. To calculate the posterior distribution, we follow the formalism described in \citet{Oppermann_2018}. We only include scans at Westerbork and Onsala L-band. Whenever Westerbork and Onsala overlap we only include scans taken with the Westerbork station to avoid possible correlations between scans (amounts to a grand total of 421.2$\,$hrs of on-source time). Therefore, we assume that all scans are independent, and calculate the total likelihood of the data as the product of the likelihoods of each individual scan. For the scan containing B1 and B2, we use the topocentric arrival time from the beginning of the scan to calculate the likelihood function. Finally, we use a uniform prior distribution and calculate the posterior distribution in the usual way as
\begin{equation}
    \text{Post}(k , r|\mathcal{D}) \propto L(\mathcal{D}|k, r)f(k, r)
\end{equation}
where $L(\mathcal{D}|k, r)$ represents the likelihood of all the data, and $f(k, r)$ represents the prior.

\subsubsection{Burst energy distribution}

To create Fig. \ref{fig:spectral_luminosity}, we made the following simplifying assumptions.  Firstly, we considered only the active phase of SGR~1935$+$2154 in April and May 2020 to estimate observing hours. During this period, for CHIME/FRB and STARE2 we assumed daily exposures of 3$\,$hrs and 9.2$\,$hrs, respectively, for 61$\,$days. FAST reported one burst in 9$\,$hrs of observing time\cite{lin_fast_2020_arxiv,fast_polarised_burst_zhang_2020ATel13699....1Z}.  For our campaign, we considered only our non-overlapping L-band observations in April and May 2020 (163.5$\,$hrs). Due to the lack of reported observed burst widths, we assume a width of $1.0\pm0.2\,$ms for the bursts reported by CHIME/FRB, STARE2 and FAST. For bursts B1 and B2 we use the values listed in Table \ref{tab:burst_properties}. Furthermore, we use the fluences as reported in the respective publications and from Table \ref{tab:burst_properties}. 

\subsubsection{Polarimetric calibration}

In our observations, we did not perform a noise-diode scan to use for polarimetric calibration. Instead, we used our test pulsar observation, PSR~J1935$+$1616, to determine the leakage correction between the recorded right and left circular polarisation (RCP and LCP). First we assume that the leakage calibration only significantly affects Stokes V (defined as $\mathrm{V = LL - RR}$, using the PSR/IEEE convention for the Stokes parameters; RR and LL are the detected power in RCP and LCP, respectively\cite{vanStraten2010}). This is approximately equivalent to moving $20\%$ of the flux density in LL to RR, i.e. we correct for a 10\%-leakage between RCP and LCP. Since Wb has an equatorial mount, we do not need to apply any corrections for parallactic angle.

\begin{figure*} 
    \resizebox{\hsize}{!}{
    \includegraphics[width=\linewidth]{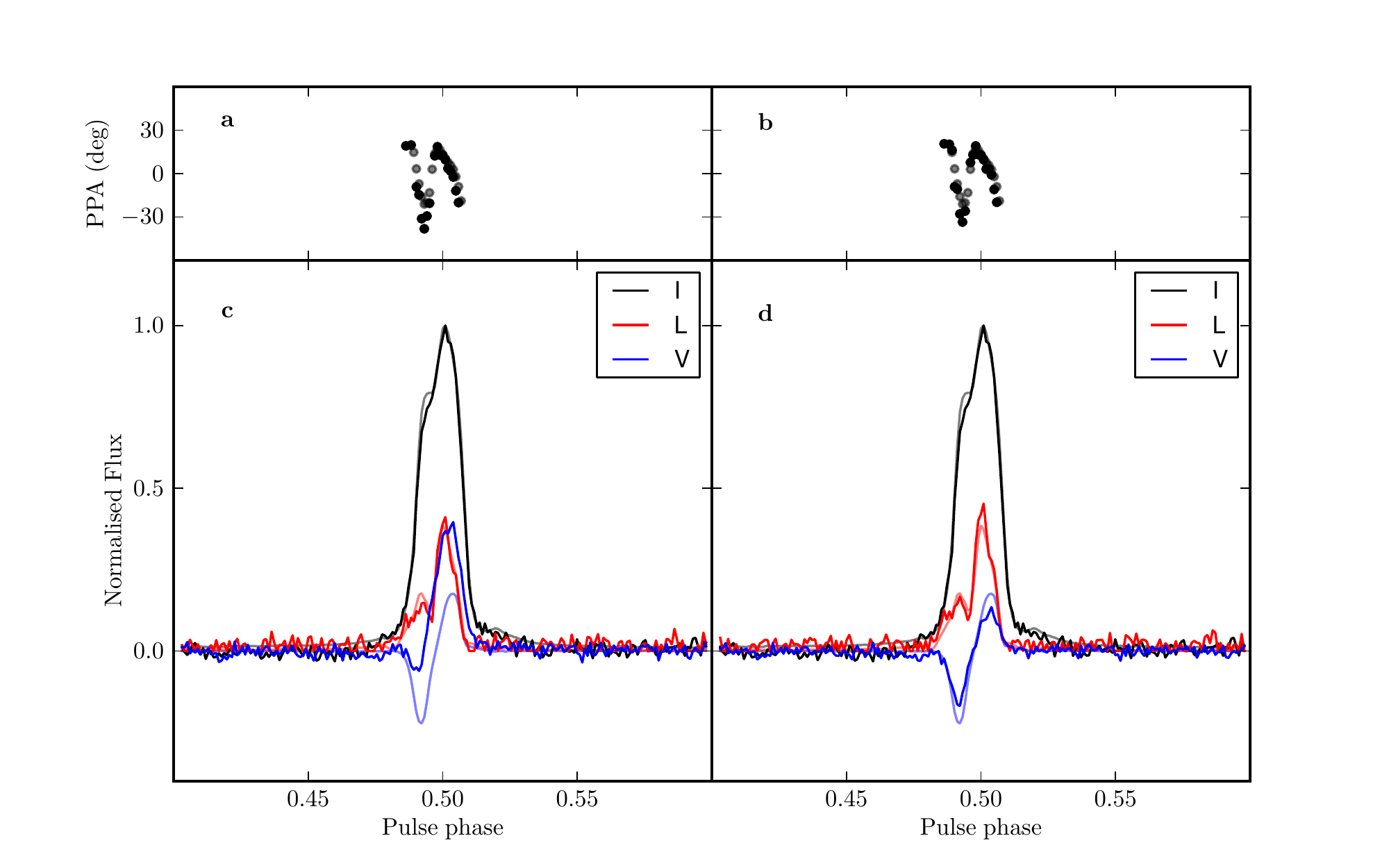}}
    \caption{The polarisation position angle swing (panels {\bf a} and {\bf b}) and average polarisation profiles (panels {\bf c} and {\bf d}) of PSR~J1935$+$1616. Shown are Stokes I (black), linear polarisation (red) and circular polarisation (blue). For comparison, the pulsar profile and PPA from the literature (at $1.4\,\mathrm{GHz}$\cite{johnston2017}) is shown using more transparent colours. {\bf a} and {\bf c}: before applying the leakage calibration discussed in Section \ref{sec:pol} and Faraday-correcting using the true rotation measure\cite{han2018} of the pulsar ($-10.2\,\mathrm{rad\,m^{-2}}$), i.e. we are also ignoring the delay between polarisation hands. {\bf b} and {\bf d}: The leakage calibrated data, Faraday-corrected using the RM determined using the PSRCHIVE tool {\tt rmfit}, which, in essence, accounts for the delay between the polarisation hands. This illustrates the polarisation calibration used for the SGR~1935$+$2154 bursts. Note that the absolute value of the PPA has been shifted to visually compare our observations with the literature. }
    \label{fig:polcal}
\end{figure*}

We still have to account for a delay between the two polarisation hands, which we assume only significantly affects Stokes Q and U. We use the tool {\tt rmfit} from PSRCHIVE, which performs a search for the rotation measure (RM) by maximising the linear polarisation fraction. Since we did not correct for the delay between the two polarisation hands beforehand, this manifests as an offset in the RM (compared with the true RM of the source), assuming the delay is constant across all frequencies. For PSR~J1935$+$1616, we measure an RM of $+77.8\,\mathrm{rad\,m^{-2}}$, which is $\sim$88 units from the true RM of $-10.2\,\mathrm{rad\,m^{-2}}$ \citep{han2018}. Under our assumption that a delay approximately corresponds to an offset in RM, an offset of $88\,\mathrm{rad\,m^{-2}}$ translates to $\sim$2$\,$ns in delay. By correcting for Faraday rotation in PSR~J1935$+$1616 using the {\tt rmfit}-determined RM ($+77.8\,\mathrm{rad\,m^{-2}}$), we reproduced the polarimetric profile and polarisation position angle (PA) swing of PSR~J1935$+$1616 within $4$\% of the published polarisation properties \citep{johnston2017}. Figure \ref{fig:polcal} shows the Faraday-corrected polarisation profile and PA swing of PSR~J1935$+$1616 using the true RM of the source, the {\tt rmfit}-determined RM and comparing both with the profile and PA presented in the literature \citep{johnston2017}.

We apply the $10\%$ leakage calibration to the bursts detected from SGR~1935+2154. We first run {\tt rmfit} to find the RM that maximises the linear polarisation. For burst B2, we find the {\tt rmfit}-measured RM to be $\sim$82$\,\mathrm{rad\,m^{-2}}$ higher than what was expected from the previously measured RM from a SGR~1935+2154 radio burst ($112.3\,\mathrm{rad\,m^{-2}}$; \citealt{fast_polarised_burst_zhang_2020ATel13699....1Z}), which is consistent with our RM offset measured for PSR~J1935$+$1616.

We then perform a joint QU fit to Stokes parameters Q/I and U/I as a function of frequency, $\nu$, using the following equations:
\begin{equation} Q/I = L\cos(2(c^2\mathrm{RM}/\nu^2 + \nu \pi D + \phi)),\end{equation}
\begin{equation} U/I = L\sin(2(c^2\mathrm{RM}/\nu^2 + \nu \pi D + \phi)),\end{equation}
where $c$ is the speed of light, and we fit for the linear polarisation fraction $L$, the delay between the hands $D$, and $\phi=\phi_{\infty}+\phi_{\rm inst}$, where $\phi_{\infty}$ is the absolute angle of the polarisation on the sky (referenced to infinite frequency), and $\phi_{\rm inst}$ is the phase difference between the polarisation hands. We perform the joint fit on Q/I and U/I spectra for PSR~J1935$+$1616 and for burst B2 from SGR~1935$+$2154, where the delay is assumed to be the same for both the pulsar scan and target scan. We fix the RM of the pulsar at the known\cite{han2018} RM of PSR~J1935$+$1616, $-10.2\,\mathrm{rad\,m^{-2}}$. We find $D\sim2.5\,\mathrm{ns}$, consistent with our prediction. Additionally, we measure the RM of B2 to be $107\pm18 \,\mathrm{rad\,m^{-2}}$, consistent with the previously measured\cite{fast_polarised_burst_zhang_2020ATel13699....1Z} value ($112.3\,\mathrm{rad\,m^{-2}}$). The fractional error on the measured RM is large since we did not perform an independent, noise-diode polarisation calibration scan, and therefore cannot remove the covariance between the fit parameters.

We debias the linear polarisation fraction following \citet{everett2001}:

\begin{equation} \label{eq:Lunbias} L_{\rm unbias}=\begin{cases}                                                                                         
    \sigma_{I}\sqrt{\left(\frac{L_{{\rm meas}}}{\sigma_{I}}\right)^2-1}, & \text{if} \ \frac{L_{{\rm meas}}}{\sigma_{I}}\ge 1.57\\                              
    0, & \text{otherwise}                                                                                                                                     
  \end{cases}  \end{equation}

where $L_{\rm meas} = \sqrt{Q^2+U^2}$, for Stokes parameters $Q$ and $U$, and $\sigma_{I}$ is the standard deviation in the off-pulse Stokes $I$.

\subsubsection{X-ray observations}

To search for X-ray bursts during the two \rm{NICER} and nine \rm{Swift}/X-ray Telescope (XRT) pointed observations, we followed standard data reduction procedures in HEASOFT v6.25 to extract light curves, using the latest calibration files via the online database \href{https://heasarc.gsfc.nasa.gov/docs/heasarc/caldb/caldb_intro.html}{{\tt caldb}}. The \rm{NICER} data were reduced using {\tt nicerdas}, applying standard filtering with additional constraints (SUN\_ANGLE $>$ $60^{\rm o}$ and COR\_SAX $>$ $4$) generated with {\tt nimaketime} and applied with {\tt niextract-events}. For \rm{Swift}/XRT, we applied the {\tt xrtpipeline} v0.13.4. After data calibration, we extracted light curves for both observatories using {\tt xselect} v2.4e at various time resolutions: $0.004$, $0.1$ and $1$ second for \rm{NICER}, $0.1$ and $1\,$s for \rm{Swift}/XRT in Window-timing mode, and $2.6\,$s for the \rm{Swift}/XRT in Photon-counting mode. Finally, we checked our methods by following the same procedures for \rm{NICER} observation 3020560101, which did not overlap with the radio campaign but was reported to contain numerous X-ray bursts\cite{younes_NICER_GBM}. We clearly recover the X-ray bursts reported therein, confirming our data reduction procedure.

For \rm{Fermi}/GBM, we focused primarily on two events: firstly, the GBM trigger on an X-ray burst of SGR~1935$+$2154\ on 2020 May 20, 21:47:07.548 UT (event bn200520908), and secondly the TTE data on 2020 May 24 22:00--23:00 UT, during which we observed radio bursts (see Section \ref{results}). For the GBM trigger data, we analysed the {\tt cspec} files of detectors n3, n6, and n7, which showed the strongest bursts in the quicklook images. Using {\tt gspec} v0.9.1, we extracted burst and background spectra per detector for the SGR~1935$+$2154 burst, which we then fitted jointly using {\tt xspec} v12.10.1. To analyse the TTE data on 2020 May 24, we used the {\tt gtbin} tool in the FERMITOOLS package to extract light curves at a $0.1$, $0.25$, and $0.004\,$s time resolutions for all twelve GBM detectors. We then used the {\tt fermi gbm data tools} v1.0.2, combined with the spacecraft pointing, to measure the viewing angle between each GBM detector and SGR~1935$+$2154. This comparison confirms that the source was visible during the radio bursts and reveals that detectors n9 and na had the smallest viewing angles, at $\sim$$41^{\rm o}$ and $\sim$$5.5^{\rm o}$, respectively.

While \rm{Fermi}/GBM triggered several additional times after the start of our radio campaign, none of these events overlapped with our radio campaign: trigger bn200503976 on 2020 May 3, also reported by \citet{ursi_may3_burst_2020GCN.27687....1U} and
\citet{li_may3_burst_2020GCN.27688....1L}, fell into a recording gap at Wb, while station O8 was affected by exceptionally strong RFI and the Tr antenna was off source. On 2020 May 10, \rm{Fermi} passed through the South Atlantic Anomaly during the X-ray burst reported by \citet{may10_burst_gcn} and no TTE data was recorded. Later on May 10, \rm{Fermi}/GBM trigger bn200510911 occurred just before the start of our radio observations.

 \subsection{Scattering time scale and scintillation}
 \citet{chime_magnetar_2020_arxiv} report a scattering time $\mathrm{\tau_{CHIME}=759\pm8\,\mu s}$ at a frequency of $600\,$MHz, while \citet{bochenek_2020_arxiv} report a scattering time $\mathrm{\tau_{STARE2}=400\pm100\,\mu s}$ at $1\,$GHz. %
 Assuming a thin screen model for scattering and  Kolgomorov turbulence, the scattering time scales with frequency as $\tau\propto\nu^{\alpha}$, with $\alpha=-4$ being the frequency scaling parameter. In this scheme, given the CHIME and STARE2 results we would expect $30\,\mu\text{s}\lesssim \tau \lesssim 120\,\mu$s at our central observing frequency $\nu=1.324\,$GHz. However, the value we measure is a factor $\gtrsim2.5$ higher ($\bar{\tau}=313\pm31\,\mu$s) and implies a frequency scaling $\alpha=-1.15$, much shallower than the canonical value. We note that the scaling implied by $\tau_{\text{CHIME}}$ and $\tau_{\text{STARE2}}$ is very similar, with $\alpha=-1.25$. Such a shallow scaling and the fact that \citet{bochenek_2020_arxiv} can reconcile their observations with no scattering, suggest that the tails that we observe could be intrinsic. Along the line of sight to SGR~1935$+$2154, the two available electron density models, NE2001\cite{cordes_2002_ne2001_2002astro.ph..7156C} and YMW16\cite{yao2017_ymw16_2017ApJ...835...29Y}, predict scattering time scales of 10$\,\mu$s and 1$\,$ms, respectively; i.e in combination they support both notions of an intrinsic tail and of a scattering tail.

A number of recent studies of pulsar scattering at low radio frequencies ($\nu<300\,$MHz) also measure values for $\alpha$ that are lower than the theoretically expected one\citep{kirsten2019ApJ...874..179K, geyer2017, meyers2017}. This can be caused by several factors among which are that the assumption of Kolmogorov turbulence and a single thin scattering screen geometry are in fact not valid. To measure the scattering time scale we assumed an intrinsic Gaussian pulse shape whose rise time can mimic that expected for an impulsive signal that travels through an extended screen, i.e. a thick screen geometry \citep{1972MNRASWilliamson}. Moreover, the assumption of a single screen might be invalid as SGR~1935$+$2154 is associated with the supernova remnant (SNR) G57.2+0.8 with high probability \citep{gaensler2014_snr_association_2014GCN.16533....1G}. Thus, besides an interstellar scattering screen about half way towards the source there could well be a second screen within the SNR, i.e. much closer to the magnetar itself. In fact, \citet{simard2020_scintillation_2020arXiv200613184S} invoke the existence of such a screen to explain the spectral structure of the burst reported by \citet{chime_magnetar_2020_arxiv}. In their model, the screen closest to the magnetar causes what can be interpreted as scintillation with a characteristic scintillation bandwidth of $\Delta\nu_{600}=100\,$MHz at an observing frequency of $600\,$MHz.  Scaled to our observing frequency this translates to $\Delta\nu_{1300}=2200\,$MHz. This is consistent with our observations in the sense that we observe during a phase of a bright scintil (caused by the screen close to the source). Any scintillation that could be caused by the interstellar screen (that is also the cause for the temporal broadening) is too narrow in bandwidth for us to resolve.

\begin{table*}
\caption{\label{tab:hxmtbursts} Times of the overlapping X-ray bursts detected by {\rm HXMT}, based on the \href{http://enghxmt.ihep.ac.cn/bfy/331.jhtml}{overview} from \citet{li_burstcatalogue_hmxt}.}
\begin{tabular}{lll}
\hline \hline
{\rm HXMT} burst time [UT] & Burst time [UT; continued] & Burst time [UT; continued] \\
\hline
2020-04-30T09:25:22.750 & 2020-05-08T06:17:16.589 & 2020-05-15T00:37:16.000 \\
2020-04-30T10:28:03.000 & 2020-05-08T09:17:05.185 & 2020-05-16T01:50:23.542 \\
2020-05-02T02:49:27.800 & 2020-05-08T09:49:21.134 & 2020-05-16T02:09:32.000 \\
2020-05-02T04:39:05.812 & 2020-05-09T01:56:38.750 & 2020-05-17T00:26:07.845 \\
2020-05-02T05:40:53.151 & 2020-05-09T23:47:15.000 & 2020-05-17T03:18:10.320 \\
2020-05-02T10:17:26.000 & 2020-05-10T01:30:01.000 & 2020-05-18T01:54:21.550 \\
2020-05-02T10:25:25.777 & 2020-05-10T01:38:45.000 & 2020-05-18T05:17:57.715 \\
2020-05-03T01:06:02.666 & 2020-05-10T03:03:38.000 & 2020-05-18T22:21:05.000 \\
2020-05-03T04:08:26.000 & 2020-05-10T03:17:15.000 & 2020-05-19T00:15:15.000 \\
2020-05-03T04:30:59.050 & 2020-05-10T05:00:28.195 & 2020-05-20T21:47:07.480 \\
2020-05-03T05:53:45.000 & 2020-05-10T06:12:01.622 & 2020-05-20T22:06:45.330 \\
2020-05-03T06:50:42.990 & 2020-05-10T06:16:41.100 & 2020-05-21T01:24:06.000 \\
2020-05-03T10:34:35.637 & 2020-05-10T06:20:09.400 & 2020-05-21T23:33:39.000 \\
2020-05-03T23:25:13.250 & 2020-05-10T06:21:26.023 & 2020-05-22T21:49:36.000 \\
2020-05-04T00:48:07.343 & 2020-05-10T06:36:51.400 & 2020-05-22T23:27:47.800 \\
2020-05-05T02:30:28.450 & 2020-05-10T08:55:46.300 & 2020-05-23T05:30:05.600 \\
2020-05-06T03:53:15.000 & 2020-05-11T02:52:18.000 & 2020-05-24T22:05:03.480 \\
2020-05-06T22:48:21.550 & 2020-05-11T04:22:52.560 & 2020-05-24T23:18:15.000 \\
2020-05-08T03:23:13.000 & 2020-05-11T23:28:40.880 & 2020-05-25T00:57:45.000 \\
2020-05-08T03:34:15.000 & 2020-05-12T06:12:09.300 & \\
\hline
\end{tabular}
\end{table*}

\newpage
\onecolumn
\begin{longtable}{lllcc}
\caption{\label{tab:observations}Details of the observations}
\\ \hline \hline MJD start$\mathrm{^{a}}$ & MJD end$\mathrm{^{a}}$ & Band$\mathrm{^{b}}$ & Time [h]$\mathrm{^{c}}$ & Station$\mathrm{^{d}}$ \\ 
\hline
\endfirsthead

\multicolumn{5}{l}{Continued from previous page} \\ \hline
MJD start$\mathrm{^{a}}$    & MJD end$\mathrm{^{a}}$   & Band$\mathrm{^{b}}$    & Time [$h$]$\mathrm{^{c}}$ &  Station$\mathrm{^{d}}$ \\ \hline 
\endhead

\hline \multicolumn{5}{l}{Continued on next page} \\
\endfoot
\hline
\endlastfoot

0.94240  & 1.28586  & C               & 7.99  & Tr \\       
0.94689  & 1.44284  & L$_{1}$         & 11.72 & O8 \\       
0.94795  & 1.22499  & P               & 4.91  & Wb \\       
1.26045  & 1.44025  & L$_{\text{Wb}}$ & 3.19  & Wb \\[1ex]  
1.94240  & 2.28586  & C               & 7.99  & Tr \\       
1.94689  & 2.44284  & L$_{1}$         & 11.72 & O8 \\       
1.94795  & 2.22499  & P               & 4.93  & Wb \\       
2.26045  & 2.43748  & L$_{\text{Wb}}$ & 3.14  & Wb \\[1ex]  
2.94240  & 3.28586  & C               & 7.99  & Tr \\       
2.94689  & 3.44284  & L$_{1}$         & 11.71 & O8 \\       
2.94795  & 3.22495  & P               & 4.94  & Wb \\       
3.26045  & 3.43470  & L$_{\text{Wb}}$ & 3.09  & Wb \\[1ex]  
3.93753  & 4.21457  & P               & 4.91  & Wb \\       
3.94241  & 4.28586  & C               & 7.99  & Tr \\       
3.94492  & 4.44087  & L$_{2}$         & 11.72 & O8 \\       
4.25005  & 4.43263  & L$_{\text{Wb}}$ & 3.23  & Wb \\[1ex]  
4.93755  & 5.21457  & P               & 4.89  & Wb \\       
4.97311  & 5.22628  & L$_{2}$         & 5.99  & O8 \\       
5.25005  & 5.42985  & L$_{\text{Wb}}$ & 3.18  & Wb \\[1ex]  
5.93753  & 6.21457  & P               & 4.93  & Wb \\       
5.94934  & 6.28679  & C               & 7.85  & Tr \\       
6.25003  & 6.42706  & L$_{\text{Wb}}$ & 3.14  & Wb \\[1ex]  
6.93752  & 7.21454  & P               & 4.92  & Wb \\       
6.93825  & 7.42361  & X               & 11.2  & O6 \\       
6.99700  & 7.22906  & L$_{3}$         & 5.49  & O8 \\       
7.25002  & 7.42428  & L$_{\text{Wb}}$ & 3.09  & Wb \\[1ex]  
7.93080  & 8.23002  & L$_{4}$         & 5.63  & O8 \\       
7.93432  & 7.98428  & P               & 1.16  & Wb \\       
8.22946  & 8.42294  & L$_{\text{Wb}}$ & 4.48  & Wb \\[1ex]  
8.92307  & 9.26666  & C               & 7.93  & Tr \\       
8.92390  & 9.09586  & L$_{\text{Wb}}$ & 3.98  & Wb \\       
8.93057  & 9.23020  & L$_{5}$         & 6.97  & O8 \\       
8.93825  & 9.42361  & X               & 11.2  & O6 \\       
9.11488  & 9.42317  & P               & 7.12  & Wb \\[1ex]  
9.92307  & 10.26666 & C               & 7.95  & Tr \\       
9.92362  & 10.39216 & L$_{5}$         & 10.97 & O8 \\       
9.92390  & 10.09586 & L$_{\text{Wb}}$ & 3.98  & Wb \\       
10.11488 & 10.41600 & P               & 6.95  & Wb \\[1ex]  
10.92307 & 11.26666 & C               & 7.95  & Tr \\       
10.92362 & 11.39218 & L$_{3}$         & 10.99 & O8 \\       
10.92391 & 11.09587 & L$_{\text{Wb}}$ & 3.98  & Wb \\       
11.11487 & 11.41600 & P               & 6.97  & Wb \\[1ex]  
11.92307 & 12.26666 & C               & 7.95  & Tr \\       
11.92362 & 12.23384 & L$_{3}$         & 7.24  & O8 \\       
11.92390 & 12.09586 & L$_{\text{Wb}}$ & 3.98  & Wb \\       
12.11488 & 12.41600 & P               & 6.96  & Wb \\[1ex]  
12.88544 & 13.34343 & L$_{3}$         & 10.74 & O8 \\       
12.92307 & 13.26666 & C               & 7.94  & Tr \\[1ex]  
14.88543 & 15.22730 & L$_{3}$         & 7.98  & O8 \\[1ex]  
15.92307 & 16.26667 & C               & 7.95  & Tr \\       
15.99524 & 16.21676 & L$_{3}$         & 5.24  & O8 \\[1ex]  
16.92307 & 17.26666 & C               & 7.95  & Tr \\[1ex]  
17.92307 & 18.26666 & C               & 7.95  & Tr \\[1ex]  
18.92307 & 19.26666 & C               & 7.95  & Tr \\[1ex]  
19.90307 & 20.06785 & L$_{\text{Wb}}$ & 3.81  & Wb \\       
19.92307 & 20.26666 & C               & 7.94  & Tr \\       
20.09058 & 20.39169 & P               & 6.96  & Wb \\[1ex]  
20.92307 & 21.26666 & C               & 7.94  & Tr \\[1ex]  
21.89622 & 22.07537 & L$_{\text{Wb}}$ & 4.14  & Wb \\       
22.09414 & 22.37373 & P               & 6.46  & Wb \\[1ex]  
22.88140 & 23.22499 & C               & 7.95  & Tr \\       
22.89623 & 23.07535 & L$_{\text{Wb}}$ & 4.14  & Wb \\       
23.09414 & 23.37373 & P               & 6.46  & Wb \\[1ex]  
23.88140 & 24.22488 & C               & 7.94  & Tr \\       
23.89623 & 24.07535 & L$_{\text{Wb}}$ & 4.14  & Wb \\       
24.09414 & 24.37373 & P               & 6.46  & Wb \\[1ex]  
24.88140 & 25.22499 & C               & 7.95  & Tr \\       
24.89623 & 25.07535 & L$_{\text{Wb}}$ & 4.14  & Wb \\       
25.09414 & 25.37373 & P               & 6.46  & Wb \\[1ex]  
25.89623 & 26.07537 & L$_{\text{Wb}}$ & 4.14  & Wb \\       
26.09414 & 26.36994 & P               & 6.24  & Wb \\[1ex]  
56.82084 & 57.24919 & L$_{3}$         & 9.49  & O8 \\[1ex]  
57.79865 & 58.26625 & L$_{\text{Wb}}$ & 9.61  & Wb \\       
57.84075 & 58.20545 & L$_{3}$         & 8.24  & O8 \\[1ex]  
58.83825 & 59.25847 & L$_{3}$         & 9.74  & O8 \\[1ex]  
59.82784 & 60.24806 & L$_{3}$         & 9.70  & O8 \\[1ex]  
60.82784 & 61.24804 & L$_{3}$         & 9.74  & O8 \\[1ex]  
61.79862 & 62.26625 & L$_{\text{Wb}}$ & 9.60  & Wb \\       
61.82784 & 62.24806 & L$_{3}$         & 9.74  & O8 \\[1ex]  
62.79170 & 63.26584 & L$_{\text{Wb}}$ & 10.44 & Wb \\[1ex]  
63.78473 & 64.25889 & L$_{\text{Wb}}$ & 10.44 & Wb \\       
63.82784 & 64.24804 & L$_{3}$         & 9.73  & O8 \\[1ex]  
64.78128 & 65.25541 & L$_{\text{Wb}}$ & 10.44 & Wb \\       
64.91228 & 65.24806 & L$_{3}$         & 7.74  & O8 \\[1ex]  
65.77779 & 66.25193 & L$_{\text{Wb}}$ & 10.44 & Wb \\       
65.82293 & 66.11082 & L$_{3}$         & 3.75  & O8 \\[1ex]  
66.77434 & 67.03318 & L$_{\text{Wb}}$ & 5.46  & Wb \\       
66.81742 & 67.09171 & L$_{3}$         & 6.49  & O8 \\[1ex]  
68.95837 & 69.23874 & L$_{\text{Wb}}$ & 5.96  & Wb \\[1ex]  
69.78128 & 70.04013 & L$_{\text{Wb}}$ & 5.47  & Wb \\[1ex]  
70.78127 & 71.23389 & L$_{\text{Wb}}$ & 9.94  & Wb \\[1ex]  
71.78128 & 72.23388 & L$_{\text{Wb}}$ & 9.94  & Wb \\[1ex]  
72.78127 & 73.23388 & L$_{\text{Wb}}$ & 9.95  & Wb \\[1ex]  
73.77087 & 74.22346 & L$_{\text{Wb}}$ & 9.65  & Wb \\[1ex]  
74.77086 & 75.22346 & L$_{\text{Wb}}$ & 9.78  & Wb \\[1ex]  
75.77087 & 76.22348 & L$_{\text{Wb}}$ & 9.94  & Wb \\[1ex]  
76.76044 & 77.21304 & L$_{\text{Wb}}$ & 9.94  & Wb \\[1ex]  
77.76046 & 78.21304 & L$_{\text{Wb}}$ & 9.92  & Wb \\[1ex]  
78.98963 & 79.20541 & L$_{\text{Wb}}$ & 4.46  & Wb \\[1ex]  
79.73963 & 80.2113  & L$_{\text{Wb}}$ & 10.37 & Wb \\[1ex]  
85.72225 & 86.19639 & L$_{\text{Wb}}$ & 10.40 & Wb \\[1ex]  
86.72226 & 87.19638 & L$_{\text{Wb}}$ & 9.13  & Wb \\[1ex]  
87.71531 & 88.18946 & L$_{\text{Wb}}$ & 10.12 & Wb \\[1ex]  
88.71530 & 89.18946 & L$_{\text{Wb}}$ & 10.45 & Wb \\[1ex]  
\hline
\multicolumn{5}{l}{$\mathrm{^{a}}$ For clarity $58968$ has been subtracted from all MJD's.} \\
\multicolumn{5}{l}{$\mathrm{^{b}}$ P:314--377$\,$MHz, L$_{\text{Wb}}$:1260--1388$\,$MHz,} \\
   \multicolumn{5}{l}{\hspace{0.3cm}L$_{1}$:1227--1739$\,$MHz, L$_{2}$:1259--1515$\,$MHz,} \\
   \multicolumn{5}{l}{\hspace{0.3cm}L$_{3}$:1360--1488$\,$MHz, L$_{4}$:1232--1488$\,$MHz,} \\
   \multicolumn{5}{l}{\hspace{0.3cm}L$_{5}$:1360--1616$\,$MHz,} \\
   \multicolumn{5}{l}{\hspace{0.3cm}C:4550--4806$\,$MHz, X:8080--8592$\,$MHz.} \\
\multicolumn{5}{l}{$\mathrm{^{c}}$ Total on source recording time in hours.} \\
\multicolumn{5}{l}{$\mathrm{^{d}}$ Wb: Westerbork RT1, Tr: Toru\'n, O8: Onsala 25m,} \\
   \multicolumn{5}{l}{\hspace{0.3cm}O6: Onsala 20m.}
\end{longtable}
\newpage
\twocolumn

\section{Data availability} 
Associated data products and plotting scripts are available at \href{https://zenodo.org/}{this placeholder link}.
The baseband data can be made available upon request to the corresponding author (franz.kirsten@chalmers.se).

\section{Code availability} 
The pipeline written to process the baseband data can be found at \url{https://github.com/pharaofranz/frb-baseband}. The code used to calculate the posterior distribution and generate Fig. \ref{fig:weibull} can be found at \url{https://github.com/MJastro95}.

\bibliography{biblio.bib}

\section{Additional information}
Correspondence and requests for material should be address to F.K.
\section{Acknowledgements}
{We would like to thank Richard Blaauw and Paul van Dijk for scheduling and running the observations at Westerbork. At Onsala we acknowledge the help of Eskil Varenius with running the observations at X-band on the Onsala 20-m telescope and we are thankful to Mitchell Mickaliger for helping with the initial setup for running FETCH. We appreciate the help of Nathalie Degenaar with interpreting the X-ray observations and we are in debt to Aard Keimpema who modified the software correlator SFXC to our needs.  We also thank George Younes for useful discussions; and we thank Shri Kulkarni, Vikram Ravi, Walid Majid, Jonathan Katz, Aaron Tohuvavohu, and Bruce Grossan for helpful comments. F.K. acknowledges support by the Swedish Research Council. J.v.d.E. is supported by the Netherlands Organisation for Scientific Research (NWO). Research at the University of Amsterdam and ASTRON is supported by an NWO Vici grant to J.W.T.H.
This work is based in part on observations carried out using the 32-m radio telescope operated by the Institute of Astronomy of the Nicolaus Copernicus University in Toru\'n (Poland) and supported by a Polish Ministry of Science and Higher Education SpUB grant.}

\section{Author contributions}
F.K. wrote and ran the search pipeline, led the observations at Onsala, interpreted the data and led the paper writing. M.P.S. ran the scattering analysis, created Figs. 1--5 and populated Table 4. M.J. performed the Weibull analysis and wrote the associated sections.  M.P.S. and M.J. co-led the Westerbork observing campaign. K.N. ran the polarisation analysis, created Fig. 6, and wrote the respective sections. J.v.d.E. searched and analysed the X-ray observations, wrote the respective sections and populated Table 3. J.W.T.H interpreted the data scientifically, supervised student work, and wrote parts of the manuscript. M.P.G. led the observations at Toru\'n and wrote the section describing those data. J.Y. supported the observations at Onsala and also interpreted the data.

\section{Competing interests}
The authors declare no competing interests.
\end{document}